\documentclass[twocolumn]{aastex63}

\shorttitle{}
\shortauthors{}
\graphicspath{{./}{figures/}}

\usepackage{amstext}
\usepackage{amsmath}
\usepackage{amssymb}
\usepackage{graphicx}

\newcommand{\beq}{\begin{equation}}
\newcommand{\eeq}{\end{equation}}
\newcommand{\beqar}{\begin{align}}
\newcommand{\eeqar}{\end{align}}

\begin{document}

\title{Tidal Wave Breaking in the Eccentric Lead-In to Mass Transfer and Common Envelope Phases}
\correspondingauthor{Morgan MacLeod}
\email{morgan.macleod@cfa.harvard.edu}

\author[0000-0002-1417-8024]{Morgan MacLeod}
\affiliation{Center for Astrophysics $\vert$ Harvard $\&$ Smithsonian 
60 Garden Street, MS-16, Cambridge, MA 02138, USA}

\author[0000-0002-3752-3038]{Michelle Vick}
\affiliation{Center for Interdisciplinary Exploration \& Research in Astrophysics (CIERA), Northwestern University, Evanston, IL 60208, USA}
\author[0000-0003-4330-287X]{Abraham Loeb}
\affiliation{Center for Astrophysics $\vert$ Harvard $\&$ Smithsonian, 
60 Garden Street, MS-16, Cambridge, MA 02138, USA}

\begin{abstract}
The evolution of many close binary and multiple star systems is defined by phases of mass exchange and interaction. As these systems evolve into contact, tidal dissipation is not always sufficient to bring them into circular, synchronous orbits. In these cases, encounters of increasing strength occur while the orbit remains eccentric. This paper focuses on the outcomes of close tidal passages in eccentric orbits.  Close eccentric passages excite dynamical oscillations about the stars' equilibrium configurations. These tidal oscillations arise from the transfer of orbital energy into oscillation mode energy. When these oscillations reach sufficient amplitude, they break near the stellar surface. The surface wave-breaking layer forms a shock-heated atmosphere that surrounds the object. The continuing oscillations in the star's interior launch shocks that dissipate into the this atmosphere, damping the tidal oscillations. We show that the rapid, nonlinear dissipation associated with the wave breaking of fundamental oscillation modes therefore comes with coupled mass loss to the wave breaking atmosphere. The mass ratio is an important characteristic that defines the relative importance of mass loss and energy dissipation and therefore determines the fate of systems evolving under the influence of nonlinear dissipation. The outcome can be rapid tidal circularization ($q\ll1$) or runaway mass exchange ($q\gg1$). 
\end{abstract}

\keywords{Hydrodynamical simulations, Tidal interaction, Close binary stars, Common envelope evolution}

\section{Introduction}

In binary and multiple star systems, the combined evolution of the stars and the orbits they share can drive systems that begin separate to directly interact. The resulting interactions include stable and unstable mass transfer, common envelope phases in which one object engulfs its companion, and complete mergers of two or more objects. Indeed, these sorts of interactions define the evolution of more than half of massive stars \citep{2012Sci...337..444S,2013ARA&A..51..269D,2017ApJS..230...15M}. In order to model how populations of multiple stars evolve and produce remnants, we need predictive theories for the mapping systems through these phases. With so many possible outcomes, performing and validating these theoretical models has been challenging and many open questions remain \citep[e.g.][]{1971ARA&A...9..183P,2002MNRAS.329..897H,2013A&ARv..21...59I,2015MNRAS.449.4415P,2021MNRAS.508.5028B}. 

The properties that describe the initial conditions for  the mapping of multiple-star systems through phases of interaction are 1) the initial state of the system orbit, and 2) the structural properties of the objects themselves.
Together, these properties define the outcome of a subsequent hydrodynamic interaction. In this paper, we focus on the property of initial orbital eccentricity at the outset of mass exchange.  Pairs of stars within multiple systems are formed with a wide distribution of initial orbital eccentricities. These initial distributions are modified by processes of tidal dissipation into the stellar interiors, gravitational torques from other bodies (e.g. secular torques in hierarchical multiples), and impulsive mass loss phases like supernovae. 

For any of these reasons, a system might evolve toward a phase of mass exchange while in an eccentric orbit \citep[e.g.][]{2016ApJ...825...70D,2016ApJ...825...71D}. In these situations, tidal dissipation competes with that evolutionary process to determine whether a mass exchange interaction begins in a circular or eccentric orbit \citep{1995A&A...296..709V,2002MNRAS.329..897H}. \citet{2020PASA...37...38V} identified, for example, that in pre-common-envelope phases involving systems that go on to produce merging double neutron star systems, the characteristic timescales for tidal circularization and spin synchronization are often longer than the radius growth timescale of the giant star in the system, apparently driving the system toward a common envelope phase while still in an eccentric orbit \citep{2016MNRAS.455.3511S,2020MNRAS.497..855S,2021MNRAS.507.2659G,2021MNRAS.508.2386S}. \citet{2021MNRAS.503.5569V} analyzed this scenario with more sophisticated models of tidal dissipation into convective stellar envelopes \citep{2020MNRAS.496.3767V} and found that the action of tides circularizes some systems, but many remain eccentric and asynchronous by the time the giant star evolves to overflow its Roche lobe at periapse. Another representative example of this class of interaction comes when secular torques drive a system toward Roche lobe overflow in an eccentric orbit, in this case tidal dissipation competes with the torque on the orbit to determine the orbital properties \citep{2014ApJ...793..137N,2016ARA&A..54..441N,2019MNRAS.487.3029S}. 

As a system evolves toward the extreme of mass transfer in an eccentric or asynchronous orbit, the primary star experiences dynamical tides of increasing strength. These tides are dynamical because they cannot be in equilibrium with the time-varying tidal potential as viewed from the stellar fluid frame. In this paper, we show that as these tides increase in amplitude and waves associated with the tidal oscillations break. In even closer passages, the amplitude becomes so large that mass is transferred from the primary to the perturber.  Wave breaking dissipates the organized motion of the oscillation into random motions, through a turbulent cascade, heat. 

Wave breaking, therefore, dissipates the energy deposited into dynamical tides when the oscillations reach an amplitude that leads them to become nonlinear. We perform hydrodynamic models of the excitation of fundamental oscillation modes by an eccentric tidal passage, and demonstrate the wave breaking that occurs when these waves reach unsustainable amplitudes. We extend our models for many dynamical times of the primary star after the periapse passage in order to quantify the rate and spatial distribution of energy dissipation associated with the damping of the fundamental mode by  wave breaking. Through this dissipation, wave breaking shapes the crucial initial conditions of system orbits and the structures of tidally-excited stars at the onset of Roche lobe overflow for eccentric systems approaching mass exchange. 

In Section \ref{sec:method}, we describe our hydrodynamic simulation methodology and the linear theory that forms a baseline for interpreting our numerical results. In Section \ref{sec:wavebreaking}, we describe the process of tidal wave breaking and the dissipation of mode energy into turbulence.  In Section \ref{sec:discussion}, we discuss some potential implications of our results, and in Section \ref{sec:conclusion}, we conclude.

\section{Methods}\label{sec:method}

\subsection{Hydrodynamic Models}

The simulations presented in this work are performed using the {\tt Athena++} software \citep{2020ApJS..249....4S}.  Our method is based on that of \citet{2018ApJ...863....5M,2018ApJ...868..136M,2019ApJ...877...28M,2020ApJ...893..106M,2020ApJ...895...29M}.  We model the interaction of a tidally-perturbed stellar envelope with two point masses, one representing the excised center of the star, and the other representing the perturbing object. Our simulations are three dimensional, and are conducted in spherical polar coordinates with their origin at the center of the star.  

The Eulerian approach of our {\tt Athena++} simulations allows high resolution in the low-mass regions at the limb of the star. This high resolution is important because it allows us to capture the shock-heating that results from wave breaking and the formation of the low-mass, high entropy atmosphere around the star. Though there are many advantages to Lagrangian hydrodynamics in stellar modeling, these mass-based methods do not have the ability to resolve the crucial features that we study in this work.

The total mass of the star is $M_1= m_1 + m_{\rm g}$, where $m_1$ is the excised core mass, and $m_{\rm g}$ is the gaseous envelope mass. The perturber has mass $M_2$, and the mass ratio is $q=M_2/M_1 = 0.1$. The original radius of the star is $R_1$. Our calculations are performed in units where $G = M_1 = R_1 = 1$, but may be rescaled to any characteristic stellar $M_1$ and $R_1$. The unit velocity is $v_1 = \sqrt{GM_1/R_1}$ and the unit time is $t_1 = \sqrt{R_1^3 / GM_1}$. 

\begin{figure}[tbp]
\begin{center}
\includegraphics[width=0.46\textwidth]{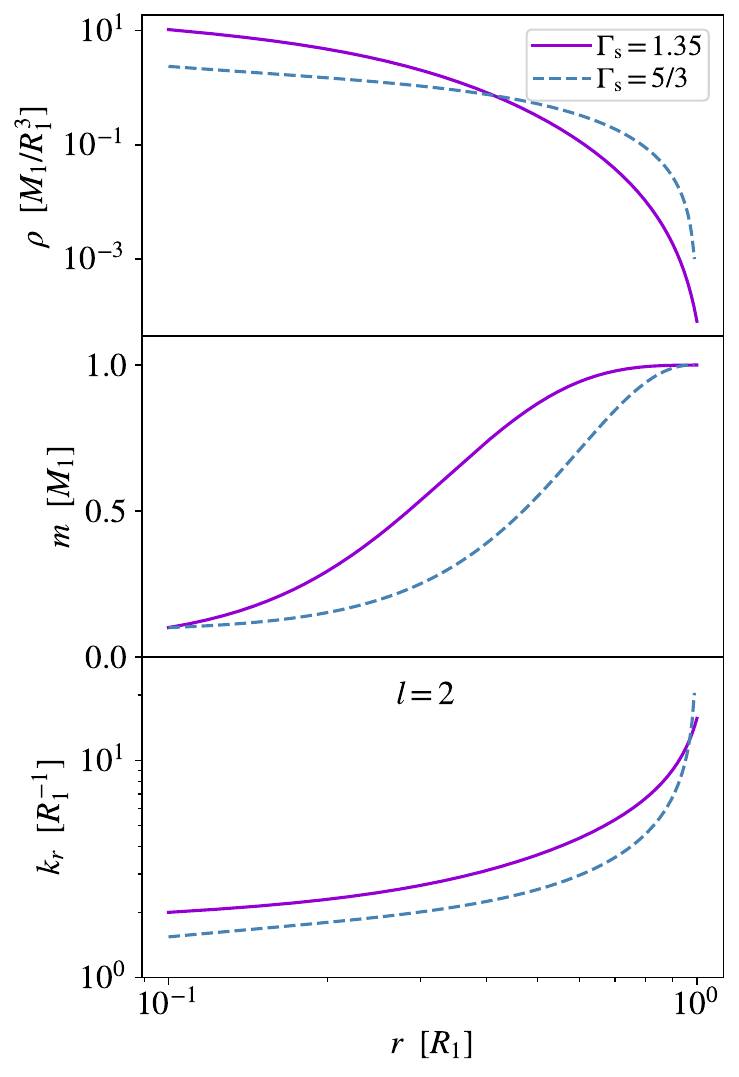}
\caption{ Structure of two model polytropic stars used in our simulations. The upper panel shows density, the central enclosed mass, and the lower the radial wavenumber $k_r \approx \omega_\alpha / c_s$, adopting the frequency of the $l=2$ fundamental mode.  Both objects have a core mass of $0.1M_1$ enclosed within $0.1R_1$, surrounded by polytropic envelopes with structural indices of $\Gamma_{\rm s}=1.35$ and $5/3$. The $\Gamma_{\rm s}=1.35$ case is more centrally-concentrated, with less mass in its outer envelope while the $\Gamma_{\rm s}=5/3$ case is more homogeneous. }
\label{fig:structure}
\end{center}
\end{figure}

The initial condition of the stellar envelope is a non-rotating spherically-symmetric polytropic envelope surrounding the excised core, which has mass $m_1 =0.1M_1$. This envelope is constructed by choosing a pressure and density at $0.1R_1$, then integrating outward to $R_1$ following the equations of of hydrostatic equilibrium closed by the polytropic index $P \propto \rho^{\Gamma_{\rm s}}$, where $\Gamma_{\rm s} = \gamma = 1.35$ or $5/3$, thus implying that the envelope is isentropic in either case. The initial density and pressure are iteratively chosen such that $M_1=1$ at $R_1=1$.  Figure \ref{fig:structure} shows the structure of our model objects. Qualitatively, these models represent giant-like stars with central condensations of mass and isentropic envelopes with differing degrees of central condensation. Outside of $R_1$, the solution is joined with a low-density hydrostatic background of constant sound speed of $ v_1/3$. 

Our frame of reference orbits with $m_1$ and is therefore accelerated, but is non-rotating. 
The coupled equations of inviscid gas dynamics that we solve are 
\begin{subequations}\label{gaseq}
\begin{align}
\partial_t \rho  + \nabla \cdot \left( \rho {\bf v} \right) &= 0 , \\
\partial_t  \left( \rho {\bf v} \right) + \nabla \cdot \left( \rho {\bf v} {\bf v} + P {\bf I} \right)  &= - \rho a_{\rm ext} , \\
\partial_t E + \nabla \cdot \left[ \left( E+ P \right) {\bf v} \right] &= - \rho a_{\rm ext} \cdot {\bf  v},
\end{align}
\end{subequations}
expressing  mass continuity, the evolution of gas momenta, and the evolution of gas energies. In the above expressions, $\rho$ is the mass density, $\rho {\bf v}$ is the momentum density, and $E = \epsilon + \rho {\bf v} \cdot {\bf v} / 2$ is the total energy density and $\epsilon$ is the internal energy density. Additionally, $P$ is the pressure, $\bf I$ is the identity tensor, and $a_{\rm ext}$ is an external acceleration (source term).  The equations above are closed by an ideal gas equation of state, $P=\left(\gamma -1\right) \epsilon$, in which $\gamma$ is the gas adiabatic index or ratio of specific heats. 

The acceleration source term on the right-hand side of equations \eqref{gaseq}, $a_{\rm ext}$, represents the forces associated with the binary, as well as fictitious accelerations associated with our choice to perform the calculation in a non-inertial frame of reference.  
The source term contains the gravitational accelerations from point masses $m_1$ and $m_2$. The acceleration of $m_2$ is softened across $0.1R_1$, using a spline kernel \citep{1989ApJS...70..419H}.  It also contains the self-gravitational acceleration of the gas on itself. This we model via a monopole approximation by summing the enclosed mass within a given radius from $m_1$, and setting the self-gravitational term acceleration equal to $-G m_{\rm g}(<r) / r^2$.  Finally, we include the inverse of the acceleration of $m_1$ to compensate for our choice of non-inertial reference frame. A more detailed account of these terms is given in equations (7)--(12) of \citet{2018ApJ...863....5M}. 

Our mesh extends over the full $4\pi$ steradians of solid angle and from $0.1R_1$ to $100R_1$. The zones are uniformly spaced in $\theta$ and $\phi$ coordinates but logarithmically spaced in $r$ so that zone shapes remain roughly cubic across a wide range of radial scales.  The base mesh is made up of 480x192x384 zones in $r$, $\theta$, and $\phi$. A nested layer of static mesh refinement increases the resolution by a factor of two between 0.7$R_1$ and $1.3R_1$. This mesh is divided into blocks of $32^3$ zones for parallelization. 

The inner radial boundary surrounding $m_1$ is located at $0.1R_1$, and is reflecting. The outer radial boundary at $100R_1$ allows outflow from the mesh but not inflow onto the mesh. The boundaries at $\theta = 0$ and $\theta = \pi$ are polar boundaries that allow material to flow through the coordinate singularity at the poles. The boundary at $\phi = 0$ and $\phi = 2\pi$ is periodic, to represent the continuous flow across this coordinate transition. Near the $\theta =0$ and $\theta = \pi$ coordinate singularities we average conserved quantities across multiple zones in the $\phi$ direction to create effectively larger, more cubical zones that avoid extreme aspect ratios near the poles. This averaging preserves the data structure of the mesh while preventing excessive $\phi$ resolution, which can severely restrict the courant-limit on the time step, in the immediate vicinity of the polar boundary. 

The orbit of the perturber and star are initialized such that the perturber follows an eccentric orbit with $a=100$ and $r_p = a(1-e)$ in the range of $1.1R_1$ to $1.9R_1$. We integrate the orbit from apoapse to a separation of $10R_1$ treating both masses as point masses, then initialize the hydrodynamic model. Particle-gas interactions are computed by direct summation over the simulation zones, and time integration is done by a leap-frog method with timestep equal to the hydrodynamic timestep.  As the model is initialized, we relax the hydrostatic profile onto the mesh, progressively turning off a damping term for $5t_1$, then turn on $m_2$ over the subsequent $0.1t_1$. The orbital integration then begins and the star proceeds toward its periapse passage with the perturber. 

\begin{figure}[tbp]
\begin{center}
\includegraphics[width=0.46\textwidth]{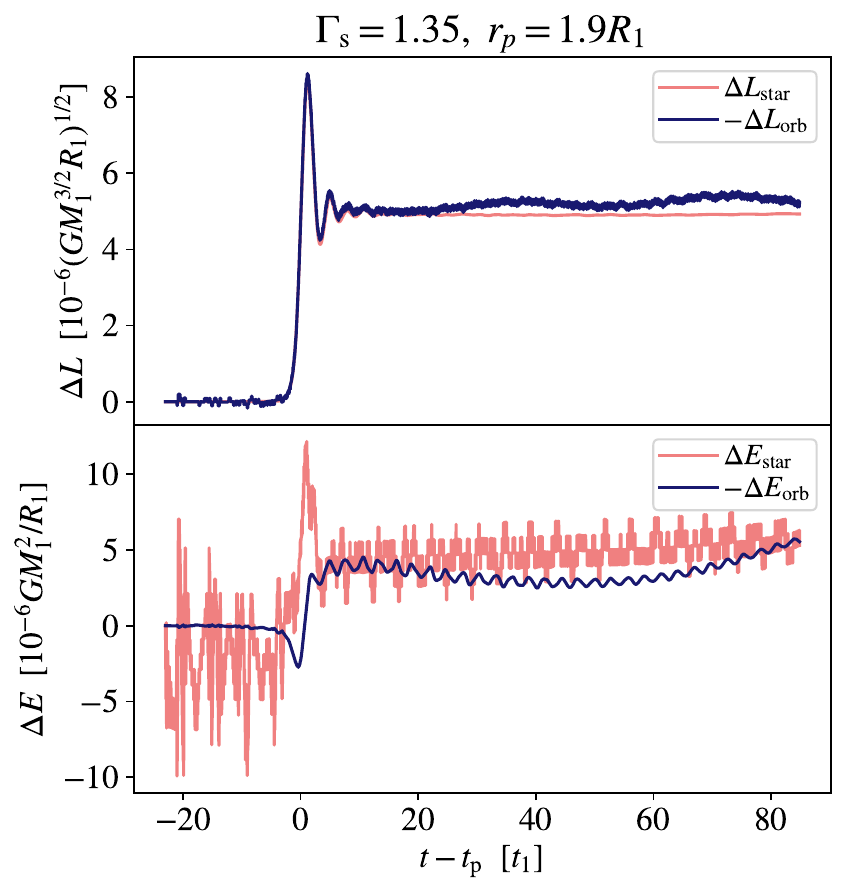}
\caption{Exchange of angular momentum and energy in the encounter between the $\Gamma_{\rm s}=1.35$ star and a perturber at $r_p=1.9R_1$ -- the weakest tidal encounter of our simulated cases. Our fiducial resolution and model parameters allow the exchange of energy and angular momentum to be accurately captured for dimensionless amplitudes $\gtrsim 10^{-6}$.  }
\label{fig:exchange}
\end{center}
\end{figure}

 At our fiducial resolution, these choices allow us sufficient fidelity to capture exchanges of energy and angular momentum with dimensionless amplitudes $\gtrsim 10^{-6}$. Figure \ref{fig:exchange} shows the exchange of angular momentum (top panel) and energy (bottom panel) between the stellar fluid and the orbit in our model. We plot this exchange in the weakest encounter for $\Gamma_{\rm s}=1.35$, which has the smallest amplitude exchange of the models we report on. Because conservation is not guaranteed in our algorithm, the departure of these quantities from perfect conservation traces the accumulation of error in our algorithm.  The overall conservation of system angular momentum in our simulations is less well preserved than the exchange, with accumulated error in the overall integration of the orbital motion accumulating into an angular momentum error at the $\sim 10^{-4}$ level after $100t_1$.  

We incorporate scalar fields that are advected with the flow in order to provide Lagrangian diagnostics of the changes in energy and momenta imparted to particular fluid parcels.  At the end of the relaxation period of damping the initial profile onto the mesh, we assign scalar tracers to represent material originating within the stellar envelope (to distinguish from the numerical background material). We also trace the initial $r_0$, $\theta_0$, $\phi_0$, coordinates, as well as the initial potential, kinetic and internal energies. These tracers then passively follow the flow, allowing us to measure Lagrangian displacements of fluid in either physical or energy space despite our Eulerian method of solving the fluid equations. 

\subsection{Linear Theory}\label{sec:lt}

We compare our numerical results with predictions for the radial velocity field from linear tidal theory (see Figure \ref{fig:spec}). In the linear theory, the response of the object to a tidal potential is a sum of the response of each eigenmode. Eigenmodes are indexed $\alpha = \{n_r,l,m\}$. We use the stellar oscillation code GYRE to calculate the $l=2-10$ f-modes ($n_r=0$) of the stellar model \citep{2013MNRAS.435.3406T}. The tidal energy transfer in a pericenter passage, $\Delta E_\alpha$, is given by equation (21) of \citet{2019MNRAS.484.5645V} assuming a non-rotating object. This calculation is similar to those of \citet{1977ApJ...213..183P} and \citet{1986ApJ...310..176L} but allows for an eccentric orbit rather than a parabolic encounter. We compute the mode masses $M_\alpha$ using equation (9) of \citet{2019ApJ...877...28M}. Equation (3) of that paper relates $\Delta E_\alpha$ and $M_\alpha$ to the radial velocity perturbation $\delta v_{\alpha}(R_1)$. The predicted power spectrum from the linear theory is then given by the sum
	\begin{equation}
		S_{ff,{\rm lt}} (l) = \frac{1}{4\pi}\sum_{m=-l}^{l} \delta v_{\alpha}^2 (R_1),
	\end{equation}
where the factor of $(4 \pi)^{-1}$ arises from the normalization of the spherical harmonics.  

Important points of reference for comparison of Lagrangian perturbations the oscillating stellar fluid  include the displacement, $\xi$, relative to the local mode wavenumber, $k$, and the velocity, $\delta v$, relative to the local sound speed.
The dispersion relation of plane acoustic waves is $k^2 = \omega^2 / c_s^2$, which relates $k$ to the oscillation frequency and local sound speed. We will focus near the stellar surface, where the radial component of $k$ is most important. We make the order of magnitude approximation that we can treat the mode near the surface as a plane wave, for which 
\beq\label{k}
k_r \sim \omega_\alpha/c_s ,   \ \ \ \ \ {\rm for} \ r \rightarrow R_1.
\eeq
Appendix E of  \citet{2010aste.book.....A} includes a more complete derivation of the local wavenumber in the asymptotic regime where the wave properties vary rapidly against the background, which is satisfied near the stellar surface. We find that equation \eqref{k} is an accurate approximation in the region of interest near the stellar limb, and adopt this simplification in what follows. The value of $k_r$, approximated by equation \eqref{k}, is plotted for our model objects in Figure \ref{fig:structure}. Although the fundamental mode has no nodes, near the stellar surface $k_r \gg R_1^{-1}$.

\section{Wave Breaking in Close, Eccentric Passages}\label{sec:wavebreaking}

In a close tidal passage, the time-dependent gravitational potential of the perturber excites tidal oscillations in the stellar fluid. In the extreme of this case, fundamental modes excited by the tide break -- implying the irreversible deformation of the wavelike motion of the oscillation -- as the perturbing force is removed and the tidal wave peak crashes back to the stellar surface.  As we discuss, the result of this irreversible deformation is dissipation. 

\subsection{Periapse Passage}

\begin{figure*}[tbp]
\begin{center}
\includegraphics[width=0.95\textwidth]{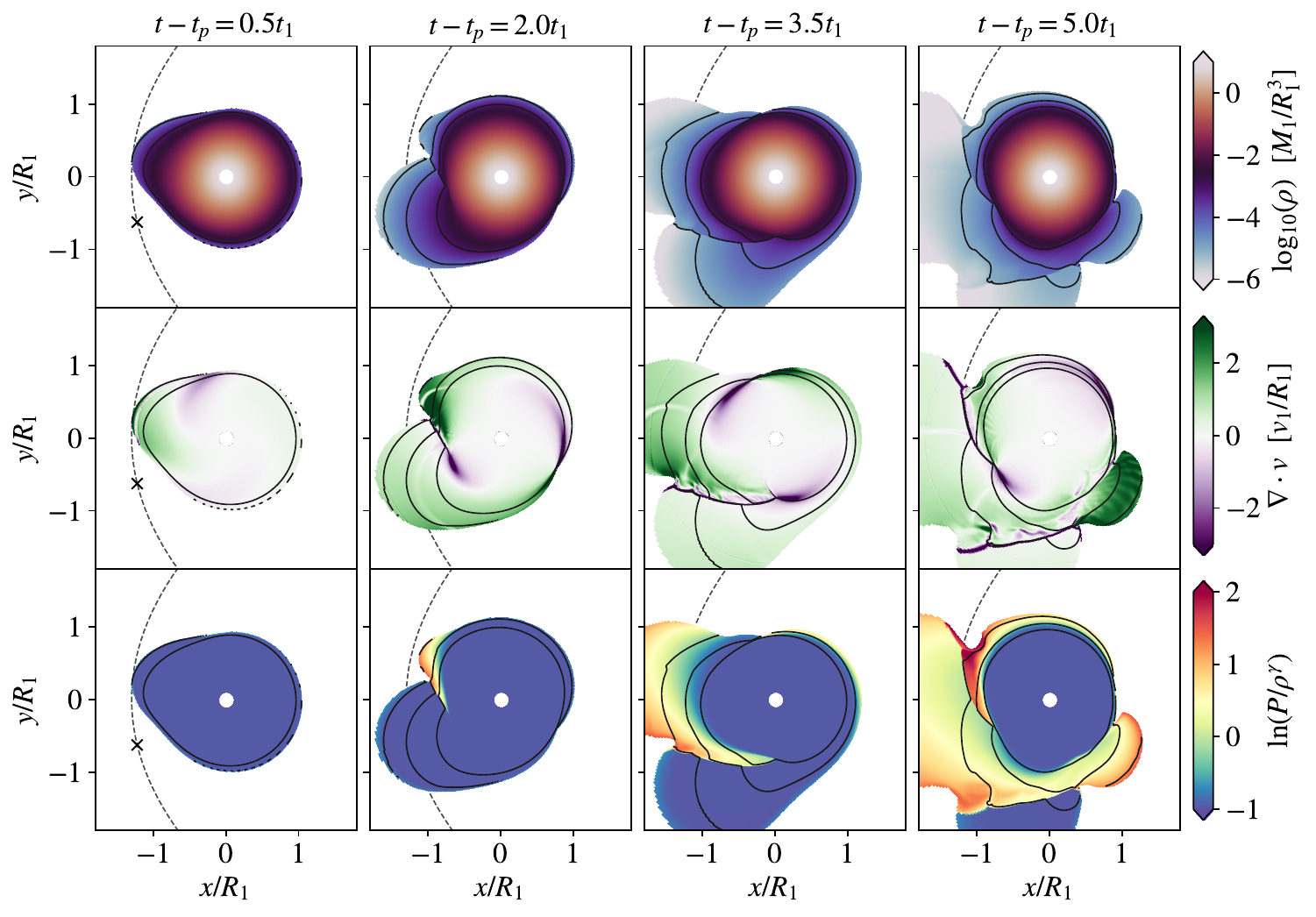}
\caption{Wave breaking during a close periapse passage. These panels show slices through the equatorial plane. Top panels show gas density, center panels show velocity divergence, and lower panels show specific entropy. Contours in each panel are at $\rho = 10^{-5}$, $10^{-4}$, and $10^{-3} M_1/R_1^3$.  The dashed line and cross marker indicate the orbital path and position of the perturbing object, respectively. In this case, the perturber passes at a periapse distance of $1.3R_1$, giving rise to a large amplitude tidal wave, which collapses on itself and breaks, forming shocks. These shocks are traced by purple regions of strong velocity convergence. Shocks increase the entropy of some of the initially isentropic stellar material.   }
\label{fig:tsp}
\end{center}
\end{figure*}

Figure \ref{fig:tsp} shows one such periapse passage with tidal wave breaking. In this case, a perturber of $M_2 = 0.1 M_1$ passes at a minimum distance $r_p=1.3R_1$. For context, this system would overflow its Roche lobe in a synchronous, circular orbit at $a\approx 1.7R_1$ \citep[In an eccentric, asynchronous orbit the critical separation for Roche lobe overflow is smaller][]{2007ApJ...660.1624S}. In the isentropic envelope, modes without radial nodes, the fundamental modes of various degree and azimuthal order, $l,m$, are the primary modes excited by a tidal passage. These modes have eigenfunctions with peak amplitude at the object's surface. 

The amplitude of the tidal wave that is excited during the periapse passage is so large that the associated radial velocities of the oscillation are greater than the local gas sound speed. With this supersonic motion, the wave loses phase coherence causing material to self-intersect and crash back to the star's surface as the perturbing force is reduced. This fallback launches shocks through the star's outer layers, highlighted by purple regions of strong velocity convergence in the center panels of Figure \ref{fig:tsp}. Their effect is to increase the specific entropy of the shocked material, traced by $\ln(P/\rho^\gamma)$ in the lower panels.

\begin{figure}[tbp]
\begin{center}
\includegraphics[width=0.47\textwidth]{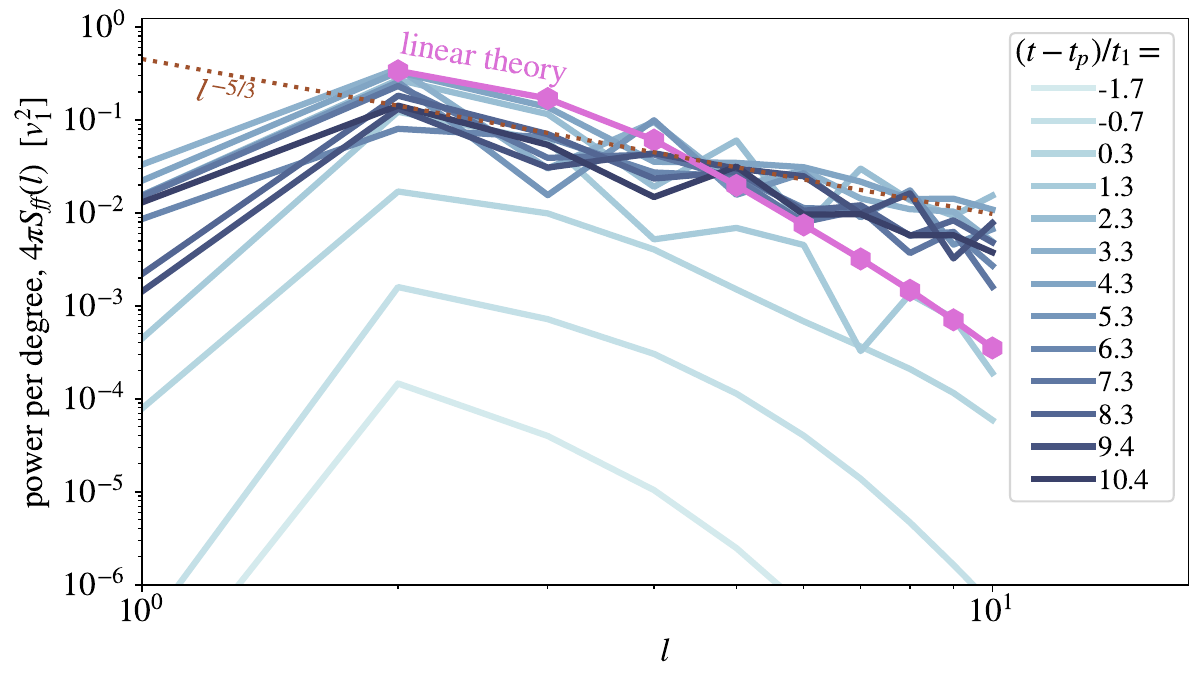}
\includegraphics[width=0.47\textwidth]{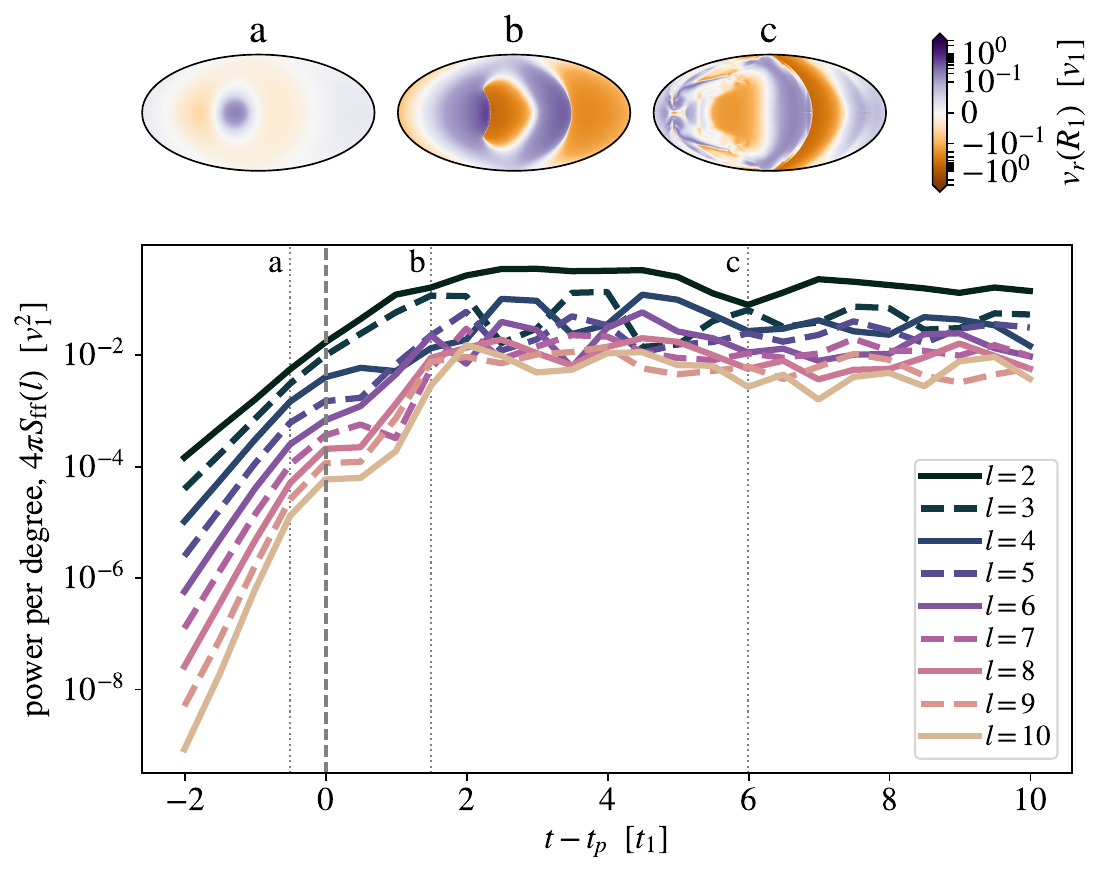}
\caption{The transformation of the spherical harmonic mode spectrum during the periapse passage due to wave breaking, in the $r_p=1.3R_1$ case. The upper panel compares the mode spectrum to that predicted by linear theory. In the lead-in to periapse, the simulated mode spectrum mirrors the linear prediction. In the immediate aftermath the mode spectrum is transformed to a power law, similar to $l^{-5/3}$, by wave breaking. In the lower panel, we see how, as the power in low-degree modes grows, wave breaking transforms the radial velocity field at $r=R_1$,  extracting power from the coherent, wave-like motion and depositing it in higher-degree modes and turbulence.   }
\label{fig:spec}
\end{center}
\end{figure}

We examine the process of fundamental mode wave breaking in this same $r_p=1.3R_1$ model in more detail by analyzing the star's fluid motion in a spherical surface at  it's original radius $r=R_1$.  Figure \ref{fig:spec} decomposes the radial velocity field  at $r=R_1$ into its spherical harmonic components. We plot the power in each angular order $l$ summed over azimuthal orders $m$,
\begin{equation}
S_{ff}(l) = \sum_{m=-l}^l  v_{lm}^2 ,  
\end{equation}
where 
\begin{equation}
v_{lm} = \frac{1}{4\pi} \int_{0}^{2\pi} \int_{0}^\pi  v_r( R_1, \theta, \phi) Y_{lm} (\theta, \phi) \sin \theta d\theta d \phi. 
\end{equation}
Figure \ref{fig:spec} shows a transformation of the power spectrum near the time of periapse, $t_p$. The initial mode spectrum, as observed from $(t-t_p)/t_1=-2$ to 0, mirrors the spectral shape predicted by the linear theory, with increasing amplitude as periapse approaches. At $(t-t_p)/t_1>1$, the spectral shape has transformed. The peak still lies in the $l=2$ quadrupole mode, but the power in higher $l$ falls off roughly as a power law. We find that the $l^{-5/3}$ scaling predicted by Kolmogorov turbulence \citep{1941DoSSR..30..301K} provides an approximate description of the mode spectrum for $l>2$. We caution, however, that our preliminary investigations have found that this power law is not universal. For example, in cases of much larger mass ratio, where the tide is very symmetric, we see a pronounced even-odd pattern in the mode power spectrum, with substantially more power in the even modes, even after wave breaking.

This transformation of the mode power spectrum by wave breaking represents the extraction of coherent energy in lower degree modes ($l=2$  and $l=3$) and its deposition into disordered motion that adds power to higher $l$ modes.  Before periapse, the energies in all of the modes are increasing, with their relative powers remaining nearly constant. Near the periapse passage, while the $l=2$ and $l=3$ power continue to rise, the $l>4$ powers momentarily flatten then dramatically increase near  $(t-t_p)/t_1=2$, as the tidal wave begins to break and shocks are launched through the material, modifying the wavelike radial velocity pattern.  The slices at $r=R_1$ of radial velocity, labeled a, b, and c in Figure \ref{fig:spec}, visualize the smoothly varying wave pattern in panel (a), the beginning of wave breaking and shocking in the abrupt transition in panel (b), and the transfer of power smaller-scale velocity features, or  higher $l, m$ modes, in panel (c).

\subsection{Conditions for Wave Breaking}

\begin{figure*}[tbp]
\begin{center}
\includegraphics[width=0.272\textwidth]{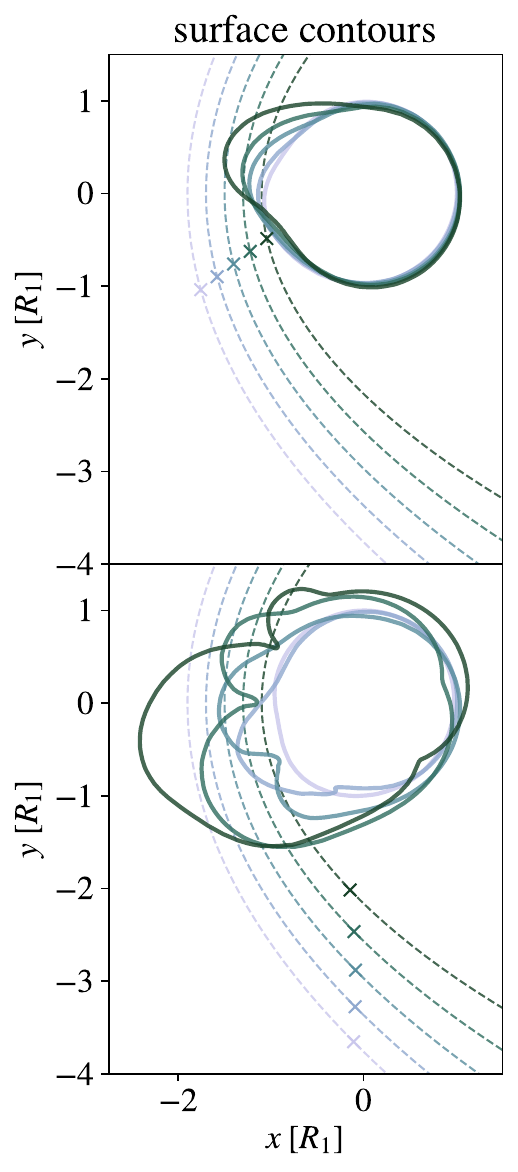}
\hspace{0.5cm}
\includegraphics[width=0.472\textwidth]{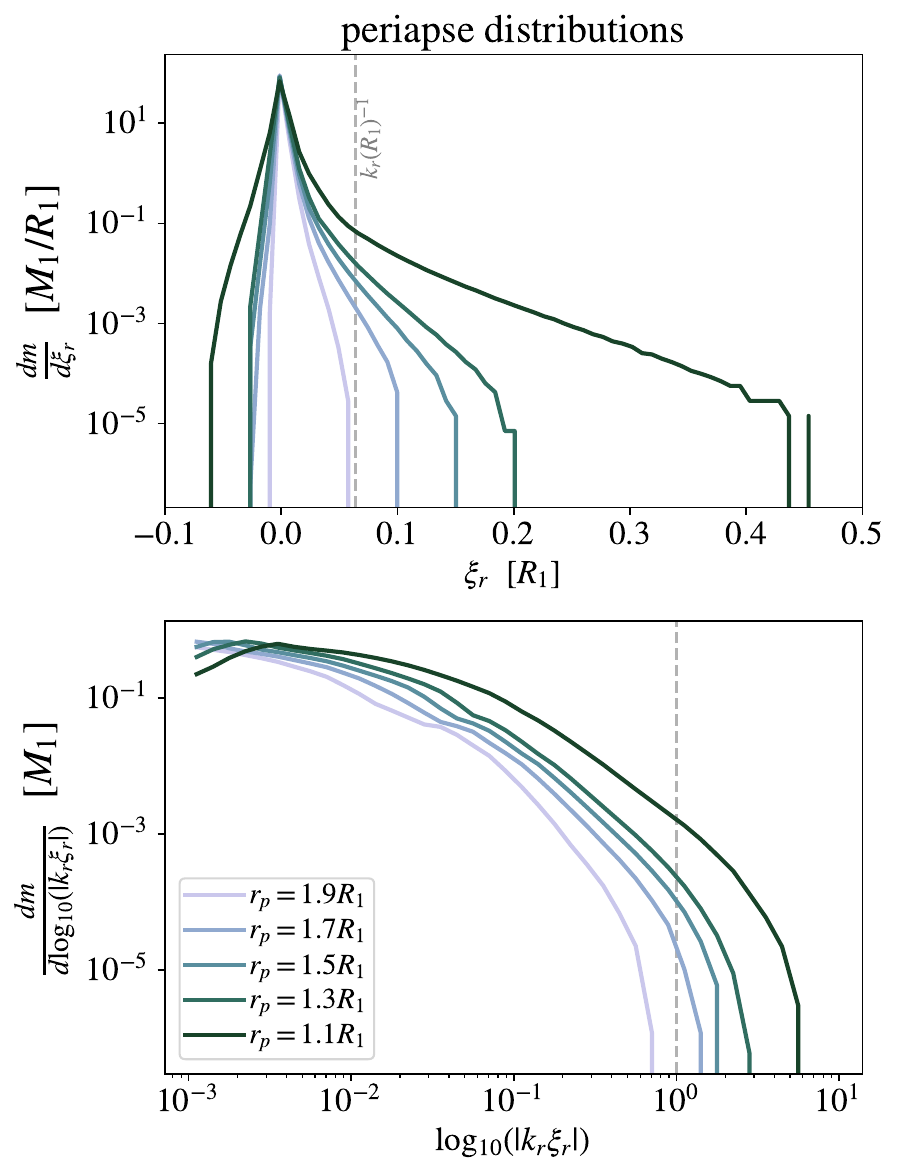}
\caption{Contours indicating the stellar surface (left panels) and mass distributions with respect to Lagrangian displacement $\xi_r$ (right panels).  Through more distorted surface contours and larger $\xi_r$, we observe that closer tidal passages lead to larger amplitude tidal waves near the star's surface. In the $r_p = 1.1 R_1$ through $r_p = 1.7$ cases, this wave breaks, and we see abrupt distortions of the wavelike profile of the surface contours. The weakest, $r_p = 1.9R_1$ case features a sufficiently small wave amplitude that it does not break, and the object surface retains its oscillatory motion. From the mass distributions, we see that the condition $| k_r \xi_r | > 1$ distinguishes the wave breaking cases from those that remain linear.   }
\label{fig:contours}
\end{center}
\end{figure*}

Figure \ref{fig:contours} explores the conditions for tidal wave breaking of the fundamental $l=2$ mode in simulations with different periapse distances from $r_p= 1.1R_1$ to $r_p = 1.9 R_1$.  The left panels show the distortion of the star's surface by the passage of the tidal perturber.  Near periapse (left panel) we see a tidally-excited wave, lagging the position of the perturbing body, that increases in amplitude the closer the periapse distance. In the second panel, taken when the orbital phase has advanced slightly, we see that the surface contours in the $r_p= 1.1$, 1.3, 1.5, and 1.7$R_1$ cases are no longer smoothly wave like, but instead host sharp reversals and multiple peaks. By comparison to the fluid properties seen in Figure \ref{fig:tsp}, we see that these surface discontinuities are the signatures of wave breaking. By contrast, the contours of the $r_p=1.9R_1$ case remain smoothly wavelike. For this particular scenario, we therefore see wave breaking occurring at periapse distances less than the separation where a synchronously-rotating star would overflow its Roche lobe in a circular orbit. 

In the right-hand panels of Figure \ref{fig:contours}, we look at the mass distributions of stellar material in radial Lagrangian perturbation, $\xi_r$, measured in snapshots that are within a half dynamical time of periapse,  $0<(t-t_p)/t_1<0.5$. Closer periapse passages yield larger stellar distortions and $\xi_r$. As described in Section \ref{sec:lt}, a point of comparison for $\xi_r$ is the radial wave number, $k_r\sim \omega_\alpha / c_s$, where the approximation is valid near the star's surface \citep{2010aste.book.....A}.  The condition $|k \xi | > 1$ is often written as an approximate criteria for wave breaking, implying that waves tend to break when their amplitudes become large relative to their scale length \citep{2010MNRAS.404.1849B,2011MNRAS.414.1365B,2020MNRAS.495.1239S}.  In our case, wave breaking is driven by the large radial displacements of the tide near the surface, so the condition becomes $|\xi_r| \gtrsim 1/k_r$ or $|k_r \xi_r| \gtrsim 1$.  The mode velocities implied are $|\delta v_r| = \omega_\alpha |\xi_r|$, which means that the the wave breaking condition can be re-expressed as a condition on supersonic radial mode velocities, $| k_r \xi_r | \approx |\delta v_r| / c_s$.  

The right hand panels of Figure \ref{fig:contours} compare $\xi_r$ to the inverse surface radial wave number, $1/k_r(R_1)$, and examine the distribution of $|k_r \xi_r| $. We see that in each of the cases where we note the presence of wave breaking ($r_p \leq 1.7 R_1$), there is material with $\xi_r > 1/k_r(R_1)$. Further, the distributions of $|k_r \xi_r| $ show that the amount of stellar mass exceeding the breaking condition of $|k_r \xi_r| > 1$ increases in stronger periapse passages.  By contrast, in the $r_p = 1.9 R_1$ case without wave breaking, $|k_r \xi_r| < 1$ for all of the star's material.  It is, however, worth noting that even when some material exceeds the breaking condition, the vast majority of the stellar material remains in the linear regime where $|k_r \xi_r| \ll 1$. For example, in the strongest encounter we simulate ($r_p = 1.1 R_1$), only $\sim 10^{-3} M_1$ is involved in the wave breaking.

\subsection{Shock-Heated Atmosphere}

\begin{figure*}[tbp]
\begin{center}
\includegraphics[width=0.95\textwidth]{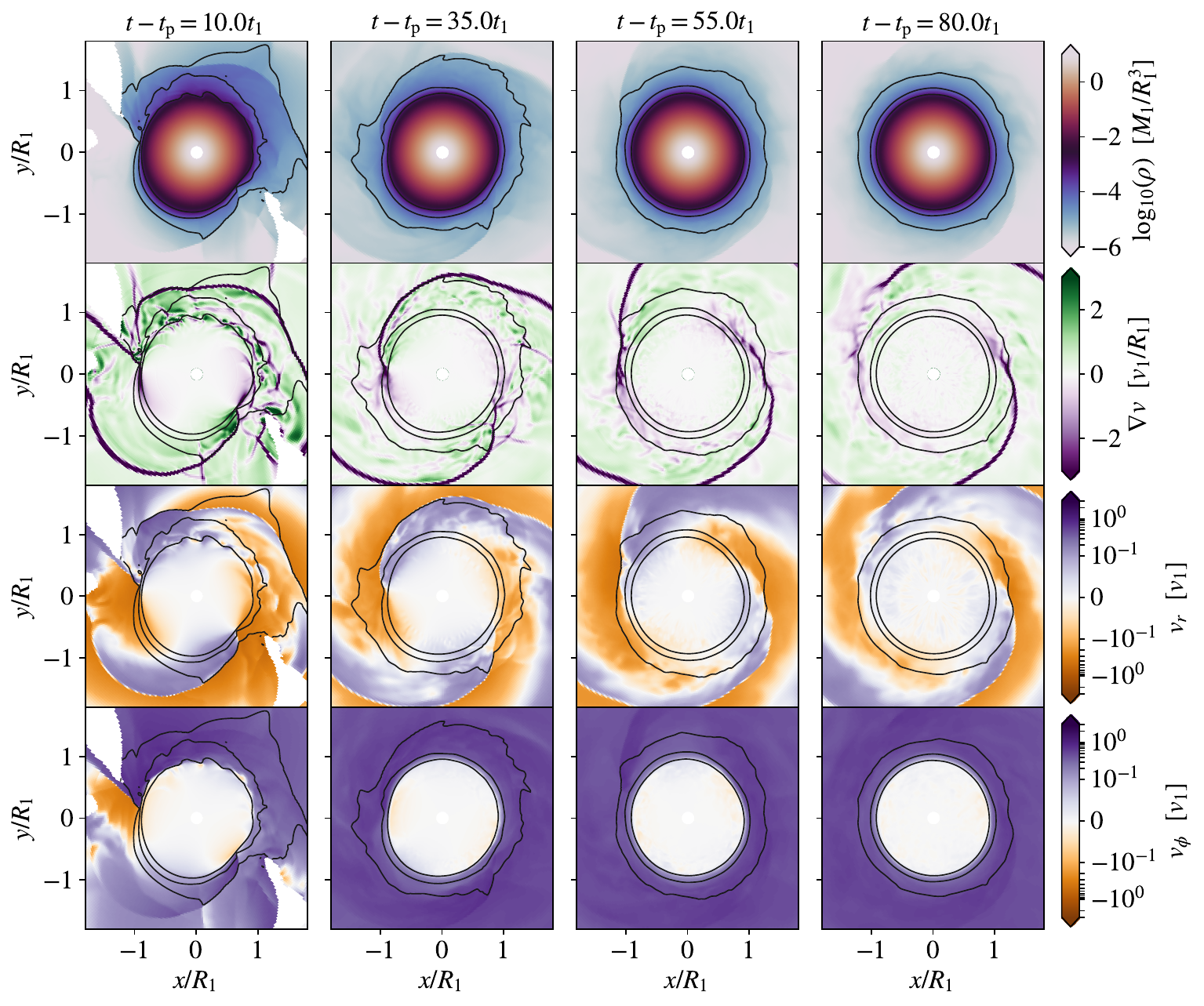}
\caption{Damping of oscillatory motions by dissipation into the turbulent atmosphere. While wave breaking during the periapse passage initially increases the specific entropy of the  surface layer, the subsequent interaction of the primary $l=2$ oscillatory mode in the  interior with this overlying atmosphere drive the damping of the oscillation, and the development of a coherently rotating surface layer subject to spiral shock waves that propagate azimuthally with the $l=2$ mode pattern speed.  }
\label{fig:tsd}
\end{center}
\end{figure*}

Just as wave breaking leads to dissipation in ocean waves, wave breaking on the surface of the star limits the maximum sustainable amplitude, $|\xi_r| \lesssim 1/ k_r(R_1)$ (or, equivelently, $|\delta v_r| \lesssim c_s (R_1)$), and dissipates tidal oscillations. The wave breaking layer damps the still-coherent linear oscillations deeper in the star's interior. The snapshots of Figure \ref{fig:tsd}  visualize the $r_p = 1.3R_1$ model across the 80 dynamical times following its periapse passage. In the leftmost panels, at 10 dynamical times after periapse passage, there are still abundant waves on the object's surface, and the surrounding atmosphere is quite chaotic. From the velocity divergence, we see that oscillations in the object's interior drive shock waves into the atmosphere. These form spiral patterns as they propagate both radially outward and azimuthally around the object with the pattern speed of the mode.  As these shocks dissipate into the atmosphere, it develops a pattern of coherent rotation at $v_\phi \sim 0.4 v_1$.

\begin{figure*}[tbp]
\begin{center}
\includegraphics[width=0.58\textwidth]{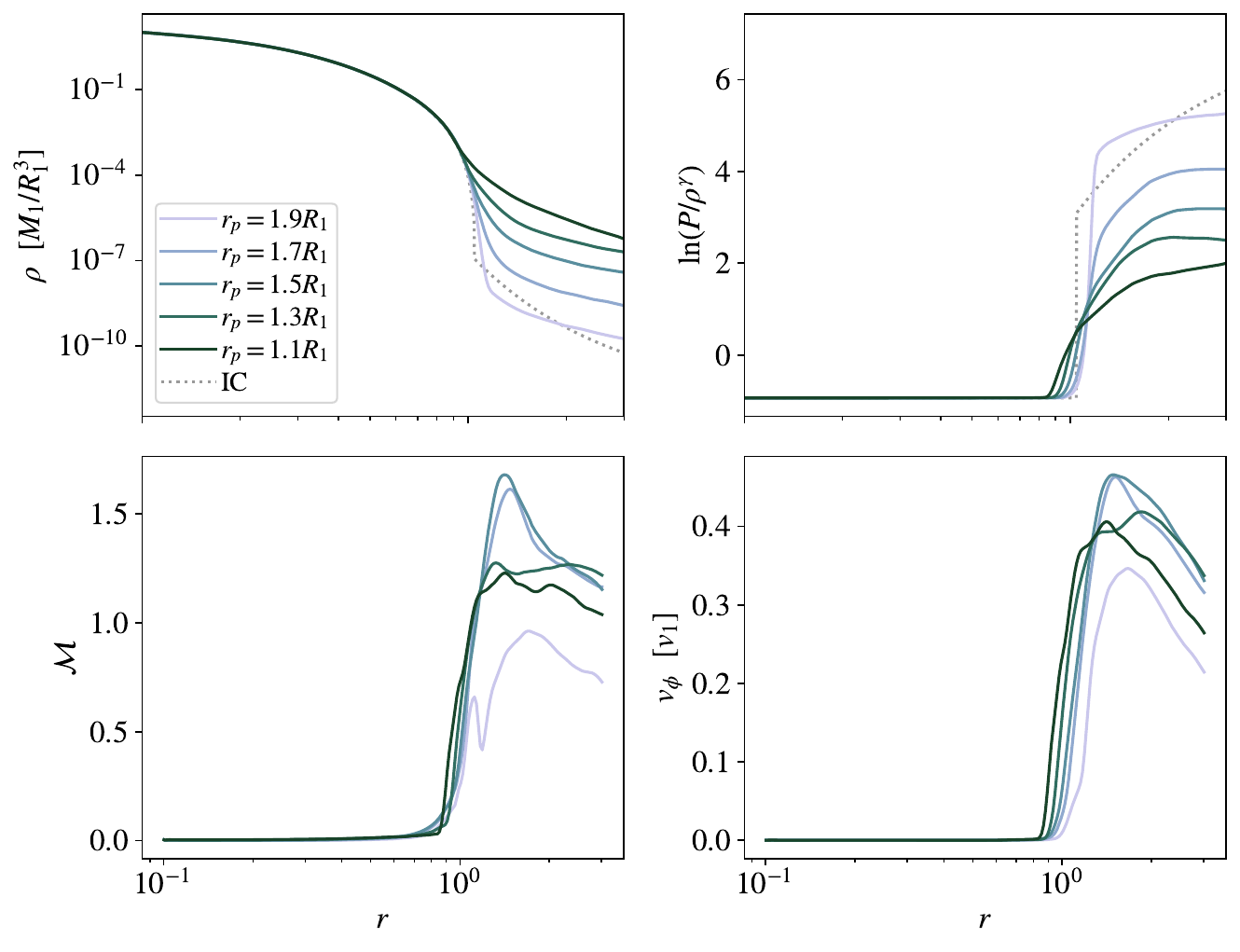}
\includegraphics[width=0.41\textwidth]{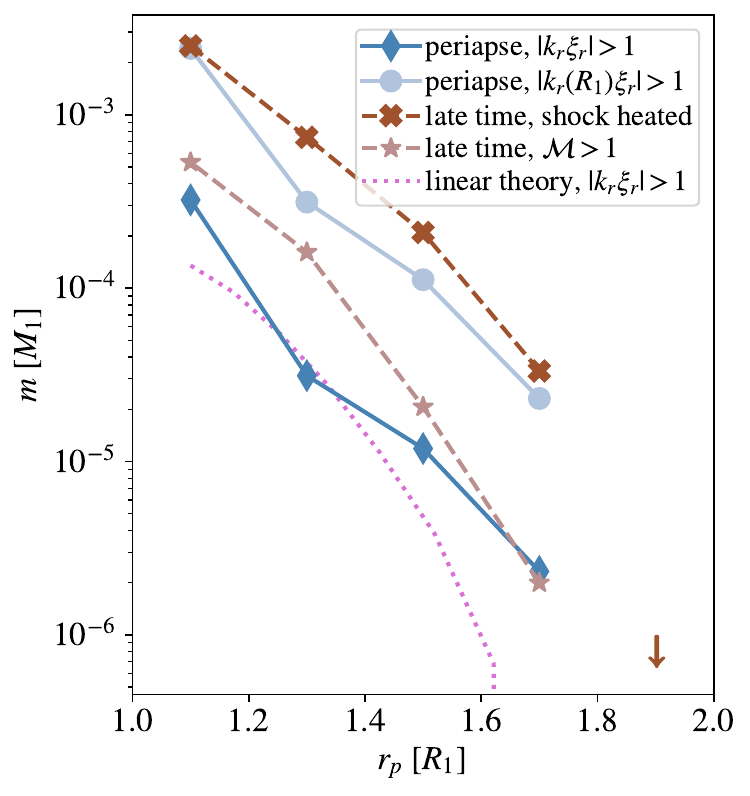}
\caption{Spherically-averaged profiles at approximately 80 dynamical times after periapse passage (left panels). We see that wave breaking has disturbed the outer layer of the object's profile adding kinetic energy (which has both rotational and turbulent components) and heat, as traced by the specific entropy.  The right panel shows the mass of the late-time atmosphere compared to the non-linear mass by several measures. We find that the linear theory prediction and simulated periapse mass with $|k_r \xi_r| > 1$ correlate with the late time mass with $\mathcal M > 1$. The total late time shock-heated mass correlates with the periapse mass with $|k_r(R_1) \xi_r| > 1$, which implies $|\delta v_r| > c_s(R_1)$.  }
\label{fig:1d}
\end{center}
\end{figure*}

The spherical averages of Figure \ref{fig:1d} exhibit the development of the rotating atmosphere layer surrounding the star, from snapshots approximately 80 dynamical times after the periapse passage. Dotted lines show the profile of the initial conditions of our simulations, which include the physical object interior and an artificial hydrostatic atmosphere that is included because vacuum conditions are not permitted in our hydrodynamic method. 
In each of the periapse passages with $r_p \leq 1.7 R_1$, an atmosphere that is significantly denser than the initial condition  develops overlying a relatively unmodified interior of the star.  Because this layer is shock-heated, it has specific entropies higher than the initially-isentropic interior of the star. The atmospheres also have similar profiles of Mach number ( $\mathcal M \gtrsim 1$) and azimuthal velocity ($v_\phi \sim 0.4 v_1$).  Meanwhile, the interior of the star maintains nearly its initial profile of density, constant specific entropy, and no coherent rotation. 

Shock waves launched by the nodes of the oscillatory waves in the star, as visualized by Figure \ref{fig:tsd}, create these shared properties of atmospheric mach number and rotation.  
The $l=2$ fundamental mode propagates around the star at an azimuthal velocity $v_\phi = \omega_2 R_1\approx 2.26 v_1$. This represents the azimuthal velocity of the shock. Taking the strong shock limit, the post shock velocity would be at least $v_\phi = \omega_2 R_1 (\gamma-1) / (\gamma + 1) \approx 0.34 v_1$.  Thus, the development of rotation in the atmosphere is related to the shock heating process and is a direct consequence of the dissipative action of shocks launched by the oscillation.  Although we have focused on the spherically-averaged properties for simplicity, this net rotation represents angular momentum in the atmosphere, which forms into a thick torus (scale height divided by radius of $h/r \sim 1$).  

The mass of the atmosphere layer can be measured by several criteria involving the transformation of properties from the interior to the atmosphere.  In the right-hand panel of Figure \ref{fig:1d}, we compare the masses exceeding the wave breaking criterion to the mass that forms the emergent atmosphere.  At late times, approximately 80 dynamical times after periapse passage, we measure the shock-heated and supersonic ($\mathcal M > 1$) atmosphere masses. The shock heated masses are determined by a specific entropy jump equivalent to a shock of $\mathcal M \geq 2$. 
These atmosphere masses are compared to masses of material that exceed wave breaking criteria at periapse, as visualized by the mass distributions of Figure \ref{fig:contours}.  We find that the non-linear mass, $|k_r \xi_r | > 1$, at periapse is very similar to the supersonic atmosphere mass at late times. The linear theory prediction of the mass with   $|k_r \xi_r | > 1$ is similar, but lower by a factor of a few, and does not predict wave breaking in the $r_p=1.7R_1$ case. This difference represents the importance of nonlinearity in these strong tidal passages. 

Meanwhile, the total shock heated atmosphere mass  at late times is more closely traced by $| k_r (R_1) \xi_r | > 1$ at periapse, which adopts the wave number at the star's surface. This criterion implies $|\delta v_r |\gtrsim c_s (R_1)$, or that oscillatory motion is more rapid than the surface sound speed. The qualitative explanation appears to be that when tidal waves break and fall back to the surface, they shock the layers of material with sound speed lower than their characteristic velocity. 

\subsection{Dissipation}

\begin{figure*}[tbp]
\begin{center}
\includegraphics[width=0.47\textwidth]{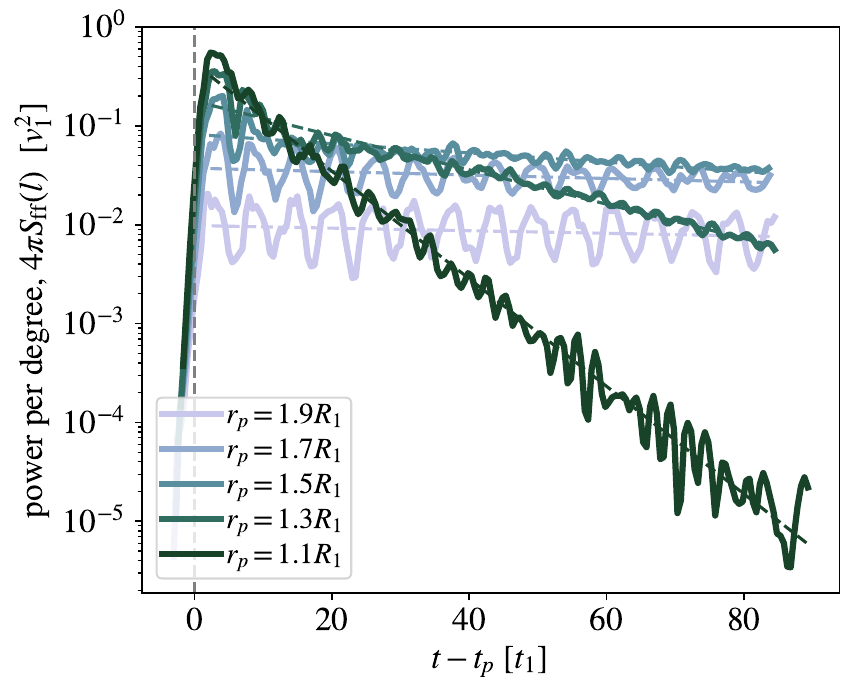}
\hspace{0.5cm}
\includegraphics[width=0.36\textwidth]{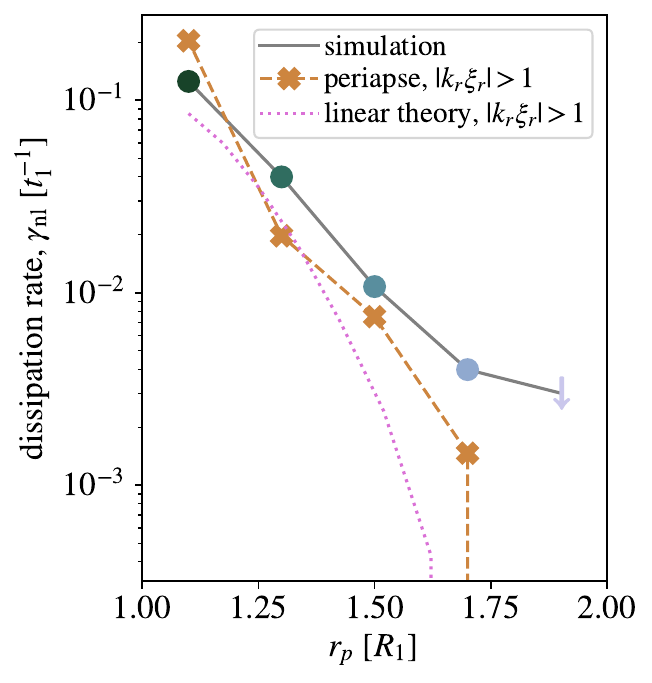}
\caption{Power in the $l=2$ mode as a function of time and its damping. After periapse passage, we see that the mode decays exponentially, with characteristic timescale that is shortest in the strongest encounters. This effect is so extreme that after $\sim 40$ dynamical times, the closest periapse passage has weaker ongoing oscillations than the furthest periapse passage. These strong encounters generate shorter dissipation timescales due to the larger masses of shocked atmosphere material generated by the initial wave breaking.  The second panel shows the dissipation rate associated with the exponential fits (the $r_p =1.9 R_1$ case is shown with a lower limit because of background material in our hydrodynamic models). The overall behavior is similar to that of our estimate of the dissipation efficiency, with $\epsilon_{\rm th} \sim 1$ and $m_{\rm atm}$ equal to the wave-breaking mass at periapse ($|k_r \xi_r| > 1$).   }
\label{fig:diss}
\end{center}
\end{figure*}

Figure \ref{fig:diss} shows the power in the $l=2$ modes over the simulation durations, evaluated at the original donor radius, $r=R_1$. After reaching a peak shortly after periapse, the mode power displays some variations about an overall decrease over time. When we measure the mode power at other locations in the star's interior, we see that it changes in proportion. Thus, the decrease over time represents a dissipation of the mode's energy, rather than a local change at $r=R_1$.  The variations come from exchanges between modes. One such example can be seen clearly in Figure \ref{fig:spec}, where at $t-t_p \approx 6 t_1$ the $l=3$ power increases as the $l=2$ power decreases, then they reverse. 

The deepest periapse passages, which experience the most extreme wave breaking, undergo the most rapid dissipation, even once their oscillatory amplitude has decreased. This means that the heirarchy of mode power by the end of the simulations is  reversed. The initially strongest tides of the $r_p = 1.1 R_1$ model have the smallest residual amplitude after 80 dynamical times. The dissipation is exponential, following a constant decay time, $\tau_{\rm diss}$. In Figure \ref{fig:diss}, we overplot best fit exponential models for the mode power for $t-t_p > 2 t_1$. In these fits,  $S_{ff}(l=2) \propto \exp \left[ - (t-t_p) \gamma_{\rm nl} \right]$.  The right hand panel of Figure \ref{fig:diss} shows that $\gamma_{\rm nl}$ is largest (most rapid dissipation) for $r_p = 1.1R_1$ and becomes progressively smaller for the larger periapse distances, following a roughly exponential trend with periapse distance.  We note here that our measurement for $r_p = 1.9 R_1$ is numerically dominated, in that it comes primarily from the background material in the initial condition of our hydrodynamic models rather than material generated by wave breaking (see the angle-averaged density in Figure \ref{fig:1d}).

Understanding how shocks dissipate energy from the oscillation into the atmosphere can lead us to a prediction for how the oscillatory modes dissipate energy over time. The oscillation energy of a fundamental mode indexed by $\alpha=\left\{l,m\right\}$ is defined as 
\begin{equation}
E_\alpha = M_\alpha \delta v_r (R_1)^2
\end{equation}
where $M_\alpha$ is the mode mass, an integral quantity of the mode eigenfunction and stellar density profile, and $ \delta v_r (R_1)$ is the characteristic velocity amplitude of the mode at the star's equator.   Because the mode amplitude and also the torus density decrease toward the poles, dissipation is equatorially concentrated. However, we might expect a prefactor of the order of unity to represent this geometric arrangement.
Therefore, in cases where wave breaking occurs, if shock waves dissipate a fraction $\epsilon_{\rm th} \sim 1$ of the specific energy of $\delta v_r (R_1)^2 / 2$, we can estimate 
\begin{equation}
\dot E_\alpha \approx {\epsilon_{\rm th} \over 2} \dot m_{\rm atm} \delta v_r (R_1)^2,
\end{equation}
where $\dot m_{\rm atm}$ is the rate that shocks sweep through atmosphere mass,
\begin{equation}
\dot m_{\rm atm} \approx \frac{m_{\rm atm}}{4\pi} m(\omega_\alpha - \Omega_{\rm atm}),
\end{equation} 
 $m$ is the azimuthal order of the mode (and thus the number of shocks launched around the circumference of the star), and $\Omega_{\rm atm}$ is the atmosphere rotation frequency. 
The characteristic decay rate of the mode due to this nonlinear dissipation, $\gamma_{\rm nl}$, can then be estimated as
\begin{equation}\label{gdiss}
\gamma_{\rm nl} \equiv -\frac{\dot E_\alpha}{2E_\alpha} \approx \epsilon_{\rm th}  \frac {m(\omega_\alpha - \Omega_{\rm atm})}{8\pi} \frac{m_{\rm atm}}{M_\alpha} .
\end{equation}
In our case, the most excited modes are the $l =2$, $m=\pm2$  fundamental modes, and $\Omega_{\rm atm} \lesssim 0.3$, so we estimate $\gamma_{\rm nl} \sim 0.15 \epsilon_{\rm th} ( m_{\rm atm} /M_\alpha )  $. 

We use our estimates of the wave-breaking mass at periapse, with $|k_r \xi_r| > 1$, as reported in Figure \ref{fig:contours}, and $M_\alpha(l=2) \approx 5.5 \times 10^{-4} M_1$, to estimate dissipation timescales.  We find that a factor $\epsilon_{\rm th} \approx 1$ provides a good fit for the normalization and overall trend of our models with differing periapse distances. Thus, damping of mode energy comes about as waves that propagate freely in the stellar interior convert to dissipative shock waves at the atmosphere boundary. The relative masses of the mode, $M_\alpha$, and the dissipative atmosphere, $m_{\rm atm} \sim m(|k_r \xi_r |_{\rm peri} > 1)$, determine the damping rate.  The predicted atmosphere masses from linear theory also exhibit a similar trend and normalization to the simulation models, but do not capture the presence of wave breaking in the $r_p = 1.7 R_1$ case.

\subsection{Dependence on Stellar Structure}\label{sec:g53}

\begin{figure}[tbp]
\begin{center}
\includegraphics[width=0.47\textwidth]{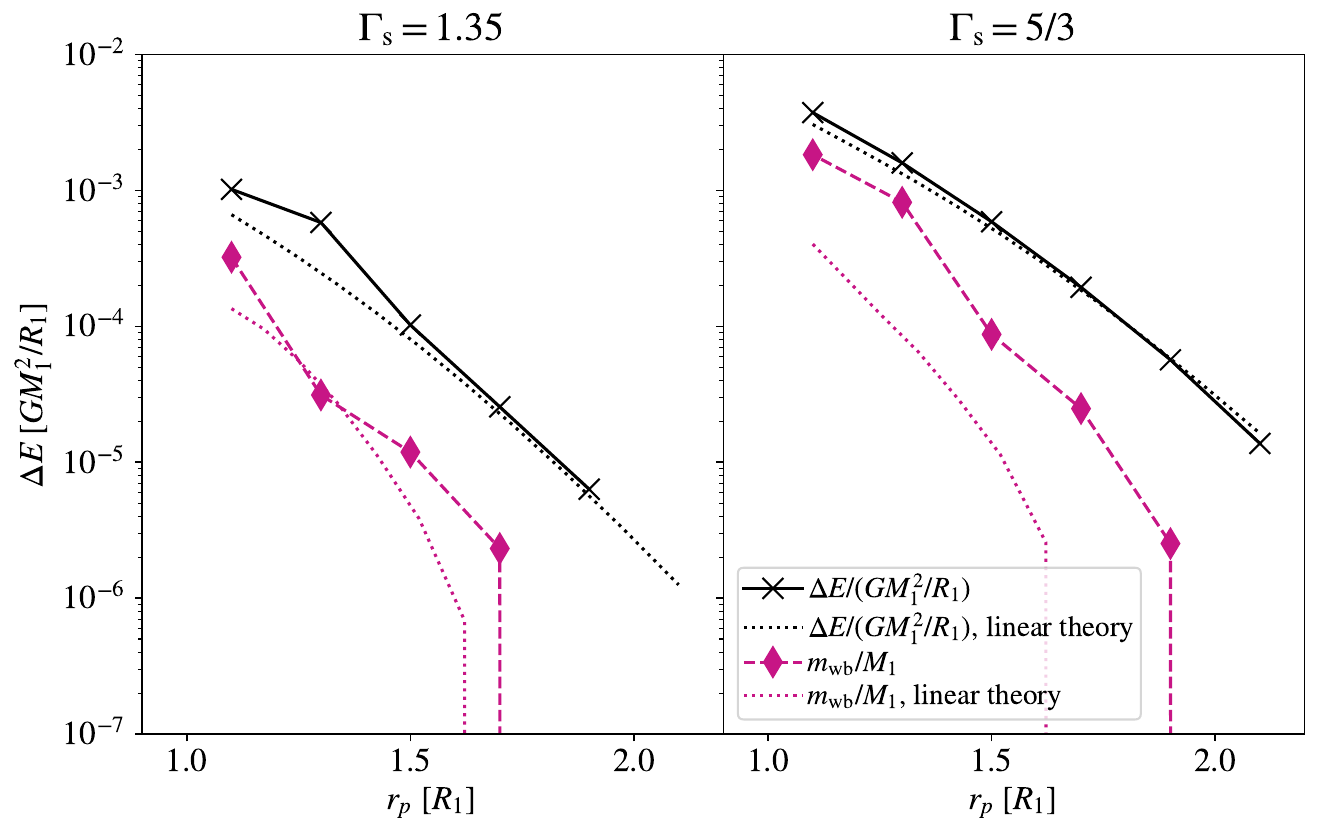}
\includegraphics[width=0.47\textwidth]{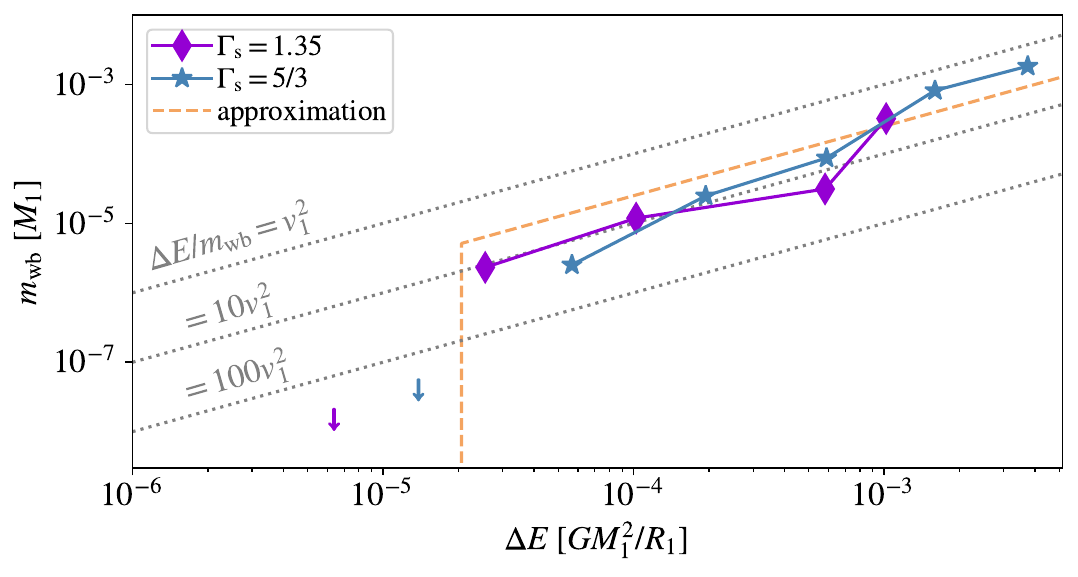}
\caption{ Simulation results for the orbital energy exchanged with tides in eccentric passages of varying periapse distance (top panel) and the wave breaking mass as a function of oscillation energy (lower panel). The dotted lines in the upper panel show the prediction of linear theory for each of the $\Gamma_{\rm s}=\gamma=1.35$ and $\Gamma_{\rm s}=\gamma = 5/3$ isentropic polytropes. The lower shows that despite differing levels of central concentration, the wave breaking mass as a function of mode energy is similar, and follows a linear relationship implying approximately constant specific energy, as traced by contours in units of $v_1^2 = G M_1/R_1$.   }
\label{fig:deltaEM}
\end{center}
\end{figure}

In this section, we briefly examine the sensitivity of our conclusions to the structure of the star, using its polytropic index as a proxy. We compare the $\Gamma_{\rm s} = 1.35$ models that have been the basis for our discussion so far to an otherwise-identical polytrope model with $\Gamma_{\rm s} = \gamma = 5/3$. This less-compressible polytrope has a less centrally concentrated density profile, and serves as a comparison to study how object structure affects the energy deposition and wave breaking mass.

We use our simulated periapse passages to measure the energy exchange between the mode and the tide along with the wave-breaking mass that results as a function of a given mode energy. These quantities are shown in Figure \ref{fig:deltaEM}. In the upper panel, we show the energy deposited in a single periapse passage, $\Delta E$, and wave-breaking mass $m_{\rm wb}$, as a function of periapse distance for each of our simulated objects. In this panel $m_{\rm wb}$ is the mass that has $|k_r \xi_r|>1$ at periapse (See Figure \ref{fig:1d}).  We compare (in dotted lines of matching color) to the linear theory predictions, summing modes from $l=2$ to $l=10$.  At a given periapse distance, more energy is deposited into oscillations of the $\Gamma_{\rm s}=5/3$ models, because these less centrally-concentrated structures have more mass in the outer layers that are most perturbed by the tidal potential. In both cases, the linear theory provides an accurate description of $\Delta E( r_{\rm p} )$.  By contrast, the linear theory underpredicts the mass in the wave-breaking layer, because this mass arises from the high-energy tail of non-linearly perturbed material, as shown in the periapse distributions of Figure \ref{fig:contours}. This is particularly evident in the more mass-rich outer layers of the $\Gamma_{\rm s} = 5/3$ model. 

The lower panel of Figure \ref{fig:deltaEM} shows the mass in the wave breaking layer as a function of energy deposited into tides. Interestingly, both polytropic structures arrive at similar results, in which we observe a roughly linear dependence between the wave breaking mass and mode energy, above a certain non-linearity threshold. This linear dependence can be cast in terms of a specific energy, as shown by the background contours in Figure \ref{fig:deltaEM}, which shows that the ratio of energy deposited over wave breaking mass is between the $GM_1/R_1$ and $10GM_1/R_1$.  This, however, does not imply that the specific energy of material in the wave breaking layer is that high, in fact it is usually at least one to two orders of magnitude lower because most of $\Delta E$ is deposited in the more mass rich lower layers at specific energies in the linear regime, and the mass exceeding the wave breaking condition represents the tail of the energy distribution. 

These results imply that there may be a relationship between energy deposited into the tidal oscillation and mass in the wave-breaking layer that does not depend on stellar structure. In the lower panel of Figure \ref{fig:deltaEM} we contextualizes this finding with a linear  approximating form, 
\beq\label{mwbapprox}
\frac{m_{\rm wb} }{M_1} \approx 
\begin{cases}
0, & {\rm for} \  \Delta E < E_{\rm nl}, \\
C_{\rm wb} \left(\frac{\Delta E}{GM_1^2/R_1} \right),  & {\rm for} \  \Delta E > E_{\rm nl},
\end{cases}
\eeq
where $C_{\rm wb}=0.25$ is a constant, and $E_{\rm nl}\approx 2\times10^{-5} GM_1^2/R_1$ is the mode saturation energy that implies wave breaking at higher energies. The upper limits in Figure \ref{fig:deltaEM} represent estimated mass-resolution limits for our simulations in which we do not observe wave breaking. In principle unresolved wave breaking could occur if the outer scale heights of the object were more finely resolved. In practice, there is likely some transition around the branches of the approximation above. An exponential cutoff of the wavebreaking mass is a physically-likely case due to the exponential density profile of the outer layers of the stellar models (and the behavoir of the linear theory predictions). However, it is not possible to constrain the shape of this cutoff with our current models, so we focus on the two clearly defined regimes of equation \eqref{mwbapprox}.

\section{Discussion}\label{sec:discussion}

\subsection{Comparison to Linear Tidal Dissipation: Location and Magnitude}

How tidal dissipation occurs, and precisely where the dissipation is located in a tidally-excited object has long been seen as a crucial question in the fate of tidally-circularizing objects \citep{1975MNRAS.172P..15F,1996MNRAS.279.1104P}. For many objects, such as giant planets around their host stars, or stars around more-massive black hole companions, the tidal object must mediate more than its own binding energy in orbital and oscillation energy exchange \citep[e.g.][]{2004MNRAS.347..437I,2016ApJ...822L..24N,2016MNRAS.458.4188M,2017MNRAS.469.3088K,2018AJ....155..118W}. In these cases, dissipation in the deep interior would quickly lead to runaway tidal inflation \citep{2001ApJ...548..466B,2003ApJ...588..509G,2004ApJ...608.1076G}. By contrast, dissipation exclusively on the surface allows for the possibility of an unperturbed interior of the object, even as large amounts of energy are exchanged \citep[e.g.][]{2018AJ....155..118W}. Our results are indicate that wave-breaking dissipation occurs exclusively in a surface layer, and further, that this overlying layer damps the oscillations of interior without heat dissipating into the interior. 

It is useful to compare these properties to various kinds of linear tidal dissipation. We have so far considered objects with convective envelopes, but when a radiative envelope is present, the principle dissipation mechanism is radiative damping of oscillations. Here radiative losses near the object surface deplete mode energy. These losses are occur because of the low optical depth in these outermost regions (though we note that this dissipation can also occur at an interior radiative--convective boundary). For low-frequency g-modes, it can be the case that $\gamma_{\rm rad} > \omega_\alpha$, which implies complete damping of the mode every cycle, and a continuous torque as it is re-excited. \citet{1975A&A....41..329Z} shows that in this limit, a crucial factor is the overlap integral between the mode and the tidal potential, which is generally very small for high-order g-modes. For fundamental modes, while the overlap integral is large,  dissipation is inefficient, such that $\gamma_{\rm rad} \ll \omega_\alpha$. One way to understand this result is that the mode eigenfunctions are such that there is significant mode energy in objects' opaque deep interiors, while sufficiently high-order g-modes might concentrate the mode energy closer to the surface \citep[e.g.][]{2012MNRAS.420.3126F}.  Thus, although radiative damping results in surface dissipation of mode energy, it is thought to be an inefficient driver of tidal orbit change in many circumstances. 

By contrast, dissipation of oscillatory motions into the random velocity field of convection can be very efficient, especially for modes that couple strongly to the tidal potential, like fundamental modes. In this case, the turbulent viscosity dissipates mode energy at a depth that depends on the mode eigenfunction, the density profile, and the turbulent viscosity as a function of radius \citep[equation 19]{1977A&A....57..383Z,2018MNRAS.481.4077S,2020MNRAS.496.3767V}.  The turbulent viscosity $\nu \sim vH$ where, $v$ and $H$ describe the characteristic velocity and length scale of eddies. Because the relevant stellar properties of the sound speed and pressure scale height decrease near a stellar surface, convective dissipation is maximized within the object interior, where radiation must diffuse outward over the thermal timescale. The diffusion time of the most dissipative layers and the adjustment of the object remain topics of intense interest \citep[e.g.][]{1996MNRAS.279.1104P,2010MNRAS.404.1849B,2011MNRAS.414.1365B,2014ARA&A..52..171O,2018AJ....155..118W,2020MNRAS.495.1239S}. Turbulent dissipation can be very strong, however, with estimated values of $\gamma_{\rm turb} \sim (M_{\rm conv}/M) (MR^2/L)^{-1/3}$. Scaled to the dynamical timescale of the object, 
\beq
\gamma_{\rm turb} \sim \left( \frac{M_{\rm conv}}{M} \right) \left( \frac{L^2 R^5}{G^3 M^5} \right)^{1/6} t_1^{-1}
\eeq
 For example, \citet{2020MNRAS.496.3767V} and \citet{2021MNRAS.503.5569V} report $\gamma_{\rm turb} \sim 10^{-5} t_1^{-1}$ for the sun's relatively low-mass mild convective layer ($\sim 0.1M_\odot$), while $\gamma_{\rm turb} \sim 0.04 t_1^{-1}$ for the sun once it ascends to the tip of the giant branch, or for the vigorous convection of near-Eddington massive stars with fully convective envelopes. 

We can therefore conclude that for fundamental modes, radiative damping is inefficient relative to nonlinear damping driven by wave breaking, and  $\gamma_{\rm rad} \ll \gamma_{\rm nl}$. However, for giants and supergiants with deep convective envelopes, it can be the case that $\gamma_{\rm turb} \sim \gamma_{\rm nl}$. This comparison carries implications for the occurrence of wave breaking and the evolution of systems under the combination of linear and nonlinear dissipation. We discuss these properties in the following two subsections.

\subsection{Conditions for Wave Breaking}
Our discussion so far has focused exclusively on our model stars, but the lessons derived can be extended to fundamental modes in a variety of stellar and planetary bodies. We find that the criterion $|k_r \xi_r| > 1$, evaluated at the object surface at periapse, reliably determines whether wave breaking occurs. By contrast, we note that wave breaking of lower-frequency $g$-modes can happen in the deep interior of an object \citep{2010MNRAS.404.1849B,2011MNRAS.414.1365B}, or at the surface \citep{2011MNRAS.412.1331F,2012MNRAS.421..426F,2012ApJ...756L..17F,2013MNRAS.430..274F,2017MNRAS.468.2296V}, depending on an object's structure. In terms of typical stellar parameters, and again adopting $k_r(R_1) \sim \omega/c_s(R_1)$, this implies wave breaking for a mode amplitude of 
\begin{equation}
\frac{|\xi_r|}{R_1} \gtrsim  10^{-2} \left( \frac{\omega_\alpha}{2  t_1^{-1}} \right)^{-1}  \left( \frac{T_{\rm eff}}{10^{-3} T_{\rm vir}} \right)^{1/2},
\end{equation}
where $T_{\rm eff}$ is the surface effective temperature and $T_{\rm vir} = (\mu m_{\rm p} / k_{\rm b}) G M_1 / R_1$ is the Virial temperature and, for the sun $T_{\rm eff} \approx 4\times 10^{-4} T_{\rm vir}$. In terms of the energy deposited by tides into a fundamental mode, 
\begin{equation}
\frac{E_{\rm nl}}{ G M_1^2 / R_1 } \sim 10^{-7} \left( \frac{M_\alpha}{10^{-3} M_1} \right)  \left( \frac{T_{\rm eff}}{10^{-4} T_{\rm vir}} \right).
\end{equation}
Where, for context, we note that the $l=2$ mode masses of our model polytropes are $5\times10^{-4}M_1$ and $10^{-2} M_1$, respectively for our $\Gamma_{\rm s} = 1.35$ and $\Gamma_{\rm s}=5/3$ cases (see Section \ref{sec:wavebreaking}). 
Thus, depositing a tiny fraction of the object's binding energy into dynamical tidal oscillations can be sufficient to drive wave breaking at the surface of realistic stars.  

Which systems evolve into wave breaking depends on their orbital and stellar evolution, which are coupled through tidal dissipation \citep[for a related discussion of dissipation and chaos, see][]{2018MNRAS.476..482V}. When $\gamma_{\rm lin} P_{\rm orb} \ll 1$, dissipation removes only a small fraction of oscillation energy each orbit and modes can accumulate over many orbital cycles. Here $\gamma_{\rm lin}$ represents the relevant linear dissipation mechanism. In this way, a star with either a radiative or small convective envelope might experience a mode energy that grows over many orbital cycles until $E_\alpha \sim E_{\rm nl}$. At that point, wave breaking occurs, and the nonlinear dissipation rate $\gamma_{\rm nl}$ becomes relevant. Because $\gamma_{\rm nl} P_{\rm orb} \gtrsim 1$ for many systems, the mode energy saturates and subsequent orbital evolution is driven by nonlinear wave breaking. By contrast, in systems with vigorous, fully-convective envelopes, $\gamma_{\rm lin} \sim \gamma_{\rm nl}$. This indicates that mode energies saturate at a level dictated by $\gamma_{\rm lin} P_{\rm orb}$ and do not necessarily grow over many orbits. This implies that when nonlinear wave breaking occurs in these systems, it is because the single periapse passage energy deposition exceeds the nonlinear wave breaking condition, $\Delta E_{\alpha} \gtrsim E_{\rm nl}$, which is the case in our hydrodynamic simulations. This scenario occurs in systems experiencing rapid stellar radius evolution, such that they remain highly eccentric while evolving toward stronger tidal passages and, eventually, mass exchange \citep{2021MNRAS.503.5569V}.

\subsection{Orbital evolution driven by nonlinear wave breaking}

As discussed in the previous subsections, wave breaking has the crucial property of dissipating heat at into the atmosphere surrounding an object, rather than its interior. This feature of wave breaking may be crucial in allowing objects to survive tidal circularization with nonlinear dissipation, even when the total energy dissipated is larger than the binding energy of the star or planet.  We find that mass loss to the wave breaking layer is an additional important process. As a system evolves through many orbits, we can imagine wave-breaking atmospheres forming when $E_\alpha \gtrsim E_{\rm nl}$. Because the atmosphere extends several times the object's radius, it seems likely that a large fraction will be stripped in the subsequent periapse passage. When nonlinear dissipation is the primary driver of orbital evolution, the rate of change of the mode energy, $\gamma_{\rm nl}$, is accompanied by a rate of change of the object mass, through the formation and subsequent removal of wave breaking layers. 

The comparison of Figure \ref{fig:deltaEM} provides crucial guidance. We observe that for two different polytropic envelope structures, there is an approximately linear relation between $E_\alpha$ and $m_{\rm wb}$, as described by equation \eqref{mwbapprox}. In the limit where nonlinear dissipation is the dominant mechanism ($\gamma_{\rm nl} \gg \gamma_{\rm lin}$) we can then express the relationship between orbital energy change and stellar mass change as 
\beq\label{dmde}
\left| \frac{d \ln M_\ast}{d \ln E_{\rm orb}} \right| \approx \frac{C_{\rm wb}}{2} q \left( \frac{R_\ast}{a} \right),
\eeq
which allows one to estimate the fractional mass-loss implied for the orbital energy to change by of order itself, and where $C_{\rm wb}\approx 0.25$ in our simulations, equation \eqref{mwbapprox}. 

The relationship between mass loss and orbital change is therefore crucial to the outcome of the nonlinear dissipation phase. For sufficiently small $C_{\rm wb}$ or $q$, equation \eqref{dmde} shows that the orbital energy can undergo large changes as the object loses small amounts of mass. If $C_{\rm wb}$ or $q$ are larger, we can have $|d\ln M_\ast / d\ln E_{\rm orb}| \gtrsim 1$, and the nonlinear phase is dominated by the mass loss from the wave breaking layer. Because many stars and planets become less dense upon mass loss, mass-loss driven evolution can lead to runaway tidal excitation and eccentric   Roche lobe overflow rather than circularization. This is precisely the behavior that was observed in the simulations of nonlinear tidal excitation on giant planets by \citet{2011ApJ...732...74G}. In these models, tides evolved to nonlinear amplitudes, through phases of fundamental mode wave breaking (see their Figure 4, for example), and into a runaway of increasing tidal excitation and mass loss \citep{2011ApJ...732...74G}.

\subsection{Radiative Cooling and Observable Signatures}\label{sec:cool}

Wave breaking converts tidal energy to heat through the disordered motions of turbulence. We can estimate the radiative diffusion times through the wave-breaking atmosphere on the basis of their optical depth, $\tau \approx \rho \kappa L$, where $L\sim R_1$ is the size scale of the atmosphere, and $\kappa$ is the opacity.  As a simple example, we take $M_1=M_\odot$, $R_1 = R_\odot$, and $\kappa \approx 0.34$~cm$^2$~g$^{-1}$. Adopting the approximation, $\rho \sim m_{\rm atm} / R_1^3$. Under these conditions, for a relatively strong encounter with $m_{\rm atm}=10^{-4}M_1$, we find $\tau \sim 10^7$, and the diffusion timescale, $t_{\rm cool} \sim \tau R_1 / c \sim 1$~yr, or approximately $10^4$ times the dynamical time of the object. The conclusion that we can draw from this estimate is that for strong encounters, radiative diffusion effectively cools atmosphere layers on timescales that are long compared to the periapse passage and subsequent mode damping, but extremely short compared to the nuclear evolution of a star.  However, a weaker encounter, such as one where wave breaking occurs only near the photosphere, where $\tau\sim1$, might experience cooling on a much shorter timescale. If cooling causes the atmosphere to reintegrate with the stellar surface layers, it could limit the dissipative efficacy of these weakest wave-breaking cases. 

The observable signatures of tidal wave breaking will similarly be varied depending on the particular objects involved along with the mass of the wave breaking layer. Wave breaking shock-heats atmosphere material to temperatures comparable to the object's Virial temperature. Thus post-shock temperatures may be of the order of $10^5$~K for a $1M_\odot$, 100$R_\odot$ giant, $10^7$~K for a main sequence star, and $10^9$~K for a white dwarf. In the presence of an optically thin atmosphere, this hot material would represent itself as an ultraviolet or x-ray excess, with luminosity matching the mode decay rate.  Scaling the example of our $r_{\rm p} = 1.7R_1$ case (Figure \ref{fig:diss}), the dissipation rate peaks at $\Delta E/ \tau_{\rm diss}$. For a sun-like star, this implies a dissipative power of $\dot E \sim  10^{38}$~erg~s$^{-1}$, near the Eddington limit luminosity and enough to drive a surface outflow \citep{2016MNRAS.458.1214Q}. Scaling to our $100R_\odot$ giant, the dissipative power is  $\dot E \sim  10^{33}$~erg~s$^{-1}$. Thus, the relative proportion of the dissipative power can represent very different fractions of the nuclear burning power of these different objects (which might be of the order of $10^{33}$~erg~s$^{-1}$ and $10^{36}$~erg~s$^{-1}$, respectively). Similarly, while the characteristic wavelength that this radiation emerges at will be dependent upon the optical depth. For example, optically thick dissipation in the presence of a dense atmosphere around a giant star could form dust and emerge as an infrared excess. 

More work is needed to clarify these many uncertainties. However, the presence of a disturbed, perhaps shock-heated surface and circumstellar material are likely to be universal -- if not unique -- traces of surface wave breaking. The case of the massive heartbeat binary system MACHO 80.7443.1718 provides one intriguing example. Here, transient emission lines disappear as the star sweeps through periapse and reappear in the aftermath \citep{2021MNRAS.506.4083J}, perhaps indicative of nonlinear tidal wave breaking in action in this system.

\section{Summary and Conclusion}\label{sec:conclusion}
We have performed simulations of close, eccentric tidal passages involving a polytropic model star and a perturber of one tenth its mass. Such a scenario might arise as a consequence of stellar evolution increasing the radius of a star in a binary system, or secular torques in a higher-order multiple system.  Indeed, recent work by \citet{2020PASA...37...38V} and \citet{2021MNRAS.503.5569V} has emphasized that, especially in systems involving massive stars, the pace of stellar evolution can outstrip that of tidal circularization implying that many such systems approach the onset of mass exchange or a common envelope phase with high eccentricities. Our simulations are relevant to what happens following that phase of gradual evolution into encounters of increasing strength. Some key findings of our study are:
\begin{enumerate}
\item In sufficiently close tidal passages, fundamental modes can become large enough in  amplitude to lose phase coherence and break, shock heating the outermost layers of the tidally-perturbed star (Figure \ref{fig:tsp}). Wave breaking dissipates the coherent energy of the oscillatory mode into the disordered motion of turbulence (Figure \ref{fig:spec}), and eventually, heat. 
\item The typical nonlinearity condition on the local wave number and amplitude, $|k_r \xi_r| \gtrsim 1$, evaluated at periapse, provides a good description of the mass involved in wave breaking (Figure \ref{fig:contours}). This mass forms a shock-heated atmosphere layer around the star (Figures \ref{fig:tsd} and \ref{fig:1d}). 
\item Ongoing oscillations in the stellar interior are damped as they steepen into spiral shock waves in the atmosphere and dissipate (Figure \ref{fig:tsd}). The nonlinear dissipation rate is set by the rate that these spiral shocks sweep through the atmosphere mass (Figure \ref{fig:diss}), and thus depends linearly on the atmosphere mass, equation \eqref{gdiss}. 
\item Compared to linear dissipation mechanisms for fundamental modes, nonlinear dissipation by wave breaking is similarly efficient to turbulent dissipation in the fully convective envelopes of giant and supergiant stars, and much more efficient than radiative dissipation. Wave breaking removes mass by lofting it into the atmosphere layer, but has the crucial property of dissipating energy only on the surface, leaving the interior unperturbed. 
\item Systems evolving primarily under the influence of nonlinear dissipation lose mass to wave breaking at a rate that is linearly proportional to the total energy change, equation \eqref{dmde}. This relationship can predict the outcome of these systems' evolution under the influence of nonlinear dissipation. When $|d\ln M_\ast / d\ln E_{\rm orb}|\ll1$, wave breaking acts to circularize the system orbit during the nonlinear phase (e.g. when $q\ll1$). When $|d\ln M_\ast / d\ln E_{\rm orb}|\gg1$, runaway tidal mass loss a likely outcome (e.g. when $q\gg1$).
\end{enumerate}
There are many directions for future study.  For example, different orbital properties or rotation of the tidal star can change the phase dependence of whether a mode propagates faster or slower than the forcing applied by a perturber. Linear theory predicts different mode spectra depending on system mass ratio, orbital eccentricity, and rotation. Although the general process of oscillation growth up to breaking at a nonlinearity threshold will hold in these systems, the particular modes and their dissipation might differ.

Because saturation at the wave-breaking limit seems to be a characteristic feature of the evolution of systems over many orbits, it would be useful to simulate the transition between the breaking and non-breaking tide with more fidelity (e.g. as shown in Figure \ref{fig:deltaEM}). Radiative cooling of the surface dissipation layers over the course of an orbital period may be important in the dissipative dynamics and fractional mass loss from the wave breaking atmosphere, especially in cases of a thin wave-breaking atmosphere near the nonlinearity limit. Relatedly, there are a broad array of astrophysical systems in which fundamental mode tidal wave breaking may shape the resultant orbital evolution. In particular, systems in which the overall change in orbital energy needed to circularize exceeds the binding energy of the objects themselves -- including eccentric formation channels for hot Jupiters and the black hole low mass x-ray binaries -- might be particularly sensitive to the details and outcome of nonlinear dissipation.

\section*{Reproduction Software and Data}

Software and data to reproduce the results of this study are publicly available in three repositories. The Athena++ simulation setup is available at \citep{morgan_macleod_2022_6325874}, post-processing and analysis software that reproduces the figures is available at \citep{morgan_macleod_2022_6577770} and selected data needed to reproduce the figures along with simulation runtime parameters are available at \citep{DVN/H5TC7Z_2022}. 

\acknowledgements
We gratefully acknowledge the feedback of the anonymous referee and many helpful discussions with colleagues that led to this study, including with J. Goldstein, J. Guillochon, D. Lai, D. Lin, E. Ostriker, P. Podsiadlowski, E. Ramirez-Ruiz, J. Stone, and Y. Wu. 
 This work was supported by the National Science Foundation under Grant No. 1909203. 
 This work used the Extreme Science and Engineering Discovery Environment (XSEDE), which is supported by National Science Foundation grant number ACI-1548562. In particular, use of XSEDE resource Stampede2 at TACC through allocation TG-AST200014 enabled this work.

\software{IPython \citep{PER-GRA:2007}; SciPy \citep{2020SciPy-NMeth};  NumPy \citep{van2011numpy};  matplotlib \citep{Hunter:2007}; Astropy \citep{2013A&A...558A..33A};  Numba \citep{numba}; Athena++ \citep{2020ApJS..249....4S}; XSEDE \citep{xsede}.  Reproduction software and data for this study \citep{morgan_macleod_2022_6325874,morgan_macleod_2022_6577770,DVN/H5TC7Z_2022}.   }

\clearpage
%\bibliography{tidaldiss,ce1,ce2,ce3,software}{}

\appendix

\section{Animated Figure}
\begin{figure}[tbp]
\begin{center}
\includegraphics[width=0.8\textwidth]{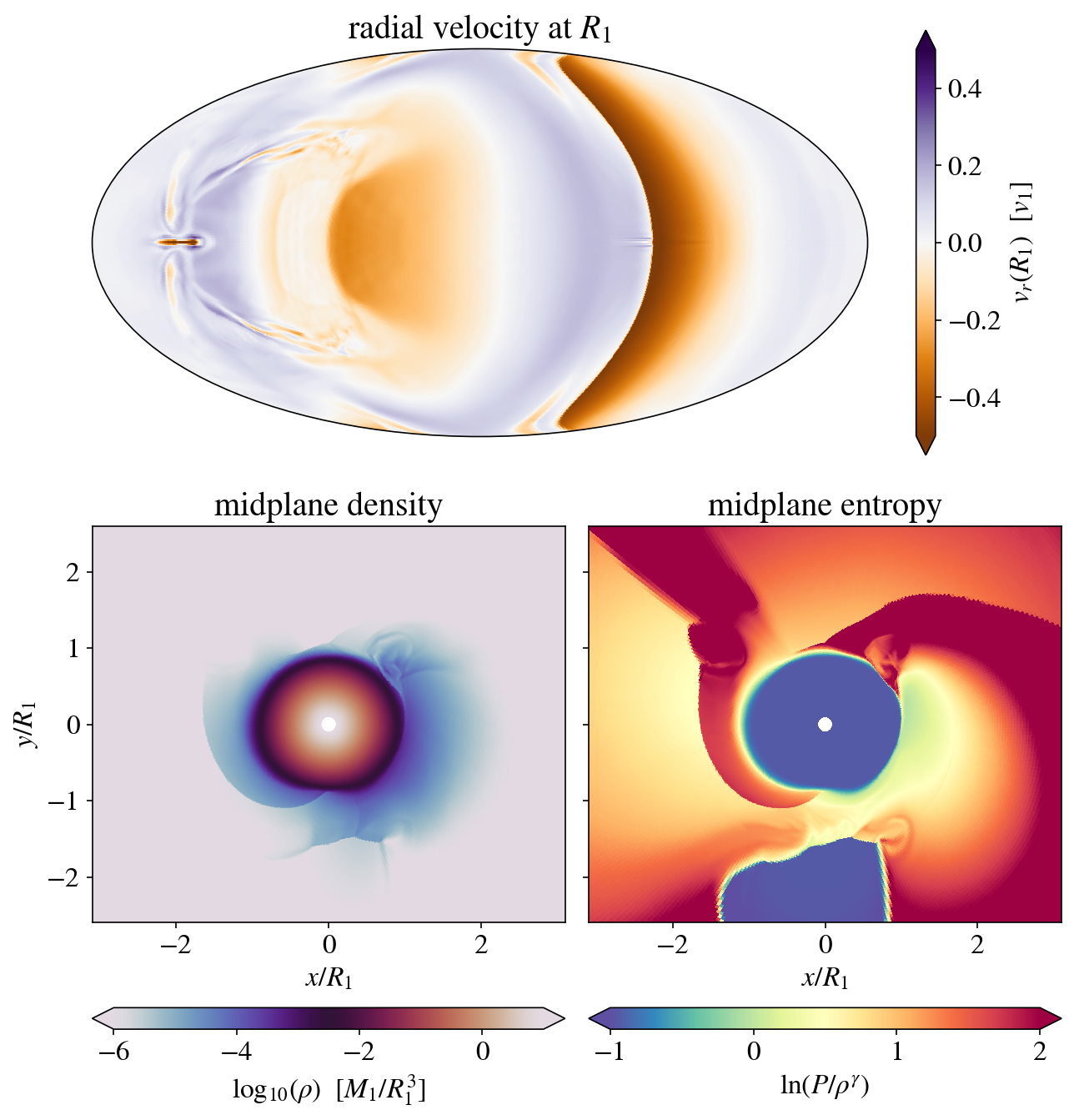}
\caption{Sample frame from an animation of the wavebreaking in the $r_{\rm p}=1.3R_1$ encounter with the $\Gamma_{\rm s}=1.35$  stellar model. }
\label{fig:anim}
\end{center}
\end{figure}

 An animation of the wave breaking encounter for $\Gamma_{\rm s}=1.35$ and $r_{\rm p}=1.3R_1$ is shown in Figure \ref{fig:anim} (animated version available in the online journal). 

\end{document}